\documentclass[11pt,a4paper]{article}
\usepackage{jheppub}

\usepackage{epsfig}

\usepackage{amsmath, amsthm, amsfonts, amssymb, latexsym}

\usepackage[caption=false]{subfig}

\usepackage{url}
\usepackage{ifpdf}

\def\beq{\begin{equation}}
\def\eeq{\end{equation}}

\def\bequ{\begin{equation}}
\def\eequ{\end{equation}}

\title{Angular correlations in TeV-gravity black hole events}

\author[a]{Marco O. P. Sampaio}
\affiliation[a]{Departamento de F\'\i sica da Universidade de Aveiro and I3N \\ 
Campus de Santiago, 3810-183 Aveiro, Portugal}

\emailAdd{msampaio@ua.pt}

 \keywords{Back Holes, Large Extra Dimensions}

\abstract{We perform a parton level study of angular correlations, in higher dimensional black hole scenarios at the LHC. We start by discussing some features of the angular spectrum in the high energy limit using the geometrical optics approximation. This allows us to recover the high energy limit of the Hawking fluxes. Then we use the full Hawking angular fluxes on the brane, for a singly rotating Myers-Perry black hole, to motivate the construction of various angular observables as to maximise the  angular asymmetries due to rotation. This is finally adapted to a parton level simulation using the \texttt{CHARYBDIS2} generator. We explore two types of variables based on: axis reconstruction and two-particle angular correlators. We find energy cuts which have the potential to help identify the effects of rotation in semi-classical rotating black hole events at the LHC during the $14~\mathrm{TeV}$ run, especially for spin-1 particles.}

\begin{document}
\maketitle




\section{Introduction}

Theories with extra dimensions have been proposed to address the hierarchy problem of the standard model of particle physics (SM)~\cite{Antoniadis:1990ew,Arkani-Hamed:1998rs,Antoniadis:1998ig,Arkani-Hamed:1998nn,Randall:1999vf,Randall:1999ee}. In such theories, the scale of gravity is lowered to the TeV energy scale and strong gravity phenomenon becomes possible, in particular black hole production as suggested by several authors~\cite{Banks:1999gd,Giddings:2001bu,Dimopoulos:2001hw,Argyres:1998qn,Eardley:2002re}. Theoretically, this is further supported by evidence for black hole formation in higher dimensions from several analytic studies~\cite{Eardley:2002re,Kohlprath:2002yh,Yoshino:2002br,Yoshino:2002tx,Yoshino:2005hi,Herdeiro:2011ck} and numerical relativity studies~\cite{Witek:2010xi,Witek:2010az,Okawa:2011fv,Zilhao:2010sr,Yoshino:2009xp}. 

From the point of view of detection, such states are predicted to decay instantaneously through Hawking evaporation~\cite{Ida:2002ez,Harris:2003eg,Harris:2005jx,Ida:2005ax,Duffy:2005ns,Casals:2005sa,Cardoso:2005vb,Cardoso:2005mh,Ida:2006tf,Casals:2006xp,Casals:2008pq} if produced at the Large Hadron Collider (LHC). Though there are still theoretical uncertainties such as the amount of gravitational radiation emitted during the collision~\cite{Yoshino:2005hi,Herdeiro:2011ck} and the corrections of the Hawking spectrum during evaporation due to gauge charges on the brane~\cite{Sampaio:2009ra,Sampaio:2009tp} or the mass of the particles~\cite{Sampaio:2009ra,Sampaio:2009tp,Herdeiro:2011uu}, current event generators already take into account the full angular spectrum with black hole rotation. Two such event generators, \texttt{CHARYBDIS2} \cite{Frost:2009cf} and \texttt{BlackMax} \cite{Dai:2007ki} are currently being used at the LHC to look for signatures of black hole production and evaporation. Bounds on these models, particularly on the fundamental planck scale $M_D$ have emerged over the years, up to the present LHC data. The main sources of bounds come from: i) deviations from Newtonian gravity in torsion balance experiments~\cite{kapner:021101,tu:201101}, ii) collider searches for Kaluza-Klein graviton production (monojet or photon with missing energy)~\cite{Aaltonen:2008hh,Aad:2011xw,Chatrchyan:2011nd} and KK graviton mediated dilepton, diphoton or dijet production~\cite{LEPexotica_graviton,Abazov:2008as,Abazov:2005tk,ATLAS:2011ab,Chatrchyan:2011fq,PhysRevLett.103.191803} iii) astrophysical or cosmological KK graviton production in supernovae and in the early universe~\cite{Hannestad:2001xi,Hannestad:2001nq}. The most restrictive laboratory bounds from current LHC searches indicate $M_D \gtrsim 2.5-3.5~\mathrm{TeV}$ for $D>6$ whereas observational bounds from astrophysics and cosmology only allow for such small $M_D$ for $D>7$. These bounds should be taken as indicative because they are very model dependent, especially the astrophysical and cosmological bounds. Regarding direct searches for black hole events at the LHC, the first attempt to establish some bounds for these models was released by the CMS~\cite{Khachatryan:2010wx} and ATLAS~\cite{Collaboration:2011bw, ATLAS-CONF-2011-068} collaborations. However, their analysis for the $7$~TeV data, depends on regions of parameter space where black holes with masses close to the unknown Planckian regime would be produced~\cite{Park:2011je}. If cuts are imposed, such that the semi-classical approximation is valid, the cross-sections become negligible at $7$~TeV, so this channel will only be properly tested in higher energy runs (originally planned to be of 14 TeV) which are likely to start by the end of 2014.

In this article, we focus on the semi-classical limit for the 14 TeV LHC runs and explore the possibility of observing angular asymmetries in the large extra dimensions scenario.  We use the default model implemented in the \texttt{CHARYBDIS2} event generator, which takes into account all of the best known theoretical calculations with black hole rotation. The purpose of this study is to focus on how to extract the physics from the events at parton level, leaving aside a detailed hadron level analysis and detector simulation.

 Our study starts with an analysis of the high energy limit of the Hawking spectrum through a geometrical optics calculation. It is well known that in the high energy limit wave propagation can be treated geometrically. Therefore, in this limit, the study of the classical trajectories of test particles outside the black hole provides useful information. The aim is to obtain the shapes of the absorptive discs as seen by an observer away from the black hole~\footnote{A first study in higher dimensions was done for some special cases in~\cite{Creek:2007sy}. We generalise their arguments to an arbitrary trajectory.} This calculation provides: i) some physical intuition which helps visualising the background and ii) the correct high energy limit of the Hawking spectrum, which could be useful to implement high energy tails in current event generators. 

Next, we provide a summary of the full spectrum of Hawking radiation on the brane, for a singly rotating Myers-Perry Black hole which is well known in the literature~\cite{Ida:2002ez,Harris:2005jx,Ida:2005ax,Duffy:2005ns,Casals:2005sa,Ida:2006tf,Casals:2006xp,Casals:2008pq,Casals:2009st}. This contains crucial information which guides our construction of various angular variables. The full parton level modelling of the events is done in the second part using the~\texttt{CHARYBDIS2} event generator which includes, in full, the effect of rotation for brane degrees of freedom.  We study two types of observables. The first type is an attempt to reconstruct the angular momentum axis of the collision. Guided by the angular distributions with the true angular momentum axis, we study a reconstruction based on the hardest particle and the sphericity tensor of the event. Though there is some correlation with the true axis we find it fails for a considerable fraction of events. The second type of variable is much more promising. We use angular correlators between pairs of particles with definite helicity. This explores the strong helicity dependence of the Hawking angular fluxes and its dependence with energy, especially at small energies (where particles with opposite helicities tend to be emitted in opposite hemispheres). We find that with suitable energy cuts, we can obtain a strong asymmetry. We suggest that the effect should survive when a full simulation is done if we explore the helicity dependence of $W$ and $Z$ decays, or use low energy jets. 

The structure of the paper is the following: In section~\ref{background} we present the background spacetime and present the geometrical optics limit calculation. We discuss how such a calculation matches well with the full Hawking spectrum and present the full Hawking spectra implemented in~\texttt{CHARYBDIS2}. In section~\ref{partonStudy} we construct several angular variables presenting an analysis of some Monte Carlo samples which show how to extract the effects of rotation at parton level. We summarise our conclusions in section~\ref{sec:Conclusions}. Some details of the calculations are left to the appendices.

\section{The Background}\label{background}
In this article we are interested in the effects of rotation on the angular emission spectrum. Though in general a brane world black hole may contain additional gauge charges (for a study see~\cite{Sampaio:2009ra,Sampaio:2009tp}) we consider the simplest case of a singly rotating Myers-Perry black hole formed on the brane with line element ($D=4+n$)
\begin{multline}\label{MP_metric}
ds^2=\left(1-\dfrac{r^2+a^2-\Delta}{\Sigma}\right) dt^2+\dfrac{\left(r^2+a^2-\Delta\right)2a\sin^2{\theta}}{\Sigma}dt d\phi-\dfrac{\Sigma}{\Delta}dr^2-\\ -\Sigma d\theta^2-\left(r^2+a^2+\dfrac{a^2 \left(r^2+a^2-\Delta\right) \sin^2{\theta}}{\Sigma}\right)\sin^2{\theta}d\phi^2-r^2\cos^2\theta d\Omega_n^2 \ , \end{multline}
where
\begin{equation}
\Delta=r^2+a^2-\dfrac{\bar{\mu}}{r^{n-1}}\,, \hspace{1cm} \Sigma=r^2+a^2\cos^2{\theta} \ .
\end{equation}
The parameters $\left\{\bar{\mu} ,a\right\}$ are related to the physical mass $M$ and angular momentum $J$ given by
\begin{eqnarray}\label{eq:MJ_MP_metric}
\dfrac{M}{M_D}&=&\dfrac{(n+2)}{2}S_{2+n}(2\pi)^{-\frac{n(n+1)}{n+2}}\,M_D^{n+1}\bar{\mu} \ , \\
J&=&S_{2+n}(2\pi)^{-\frac{n(n+1)}{n+2}}\,M_D^{n+2}\,a\,\bar{\mu} =\dfrac{2}{n+2}\,M a \ ,
\end{eqnarray}
where $M_D$ is the higher dimensional Planck mass and $S_{2+n}$ is the area of the $n+2$-sphere. The horizon radius and oblateness $\left\{r_H,a_*\right\}$ are defined through the largest positive root of $\Delta(r_H)=0$ ($r_H$ is directly related to the surface curvature of the horizon) and $a_* = a/r_H$ (which is the oblateness of the spheroidal horizon) respectively.

The theory of Hawking radiation predicts that black holes emit a continuous flux of particles. In horizon radius units ($r_H=1$), the differential fluxes of particle number, energy and angular momentum are~\cite{Hawking:1974sw}
\begin{equation}
\frac{d \left\{N,E,J\right\}_s}{dt d\omega d\Omega} = \frac{1}{2\pi} \sum_{j=|s|}^\infty \sum_{m = -j}^{j} \frac{\left\{1,\omega,m\right\}}{\exp(\tilde{\omega}/T_H) -(-1)^{2|s|}} \mathbb{T}^{(4+n)}_{s}(\omega, a)\left|{}_{s}S\left(a\omega,\cos\theta\right)\right|^2\;,  \label{eq-flux-spectrum}
\end{equation}
where $\tilde{\omega}=\omega-m\Omega_H$, $k=\{j,m\}$ are the angular momentum quantum numbers, $h=-s$ is the helicity of the particle,
\begin{equation}
T_H=\dfrac{(n+1)+(n-1)a^2}{4\pi(1+a^2)r_H}
\end{equation}
and $\Omega_H=a/(1+a^2)$ is the angular velocity of the horizon. $\mathbb{T}^{(4+n)}_k$ are the so called transmission factors defined as the fraction of an incident wave from infinity which is absorbed by the black hole. The boundary conditions are such that close to the horizon the wave is purely ingoing for co-rotating physical observers~\cite{Bardeen:1972fi}. The term in denominator, is usually called the Planckian (or thermal) factor. In the next sub-sections we analyse some of the properties of the Hawking radiation that are expected in this background.

\subsection{The high energy limit}

\begin{figure}[t]
\centering\includegraphics[scale=0.4,trim= 0cm 0 0cm 0cm,clip=true]{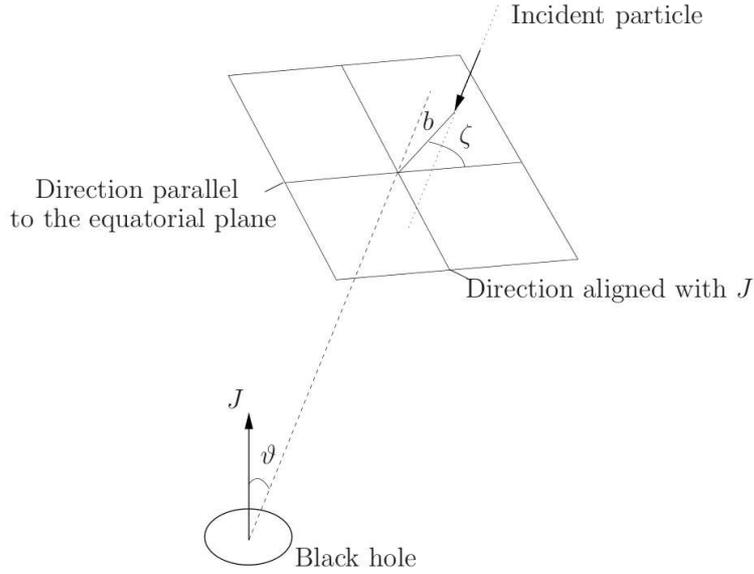}
\caption{\label{fig:diagram_zeta} Diagram defining the $\zeta$ angle and the impact parameter $b$ on the plane transverse to the direction of incidence.}
\end{figure}

We start with a study of the geometrical optics (or test particle) limit. We focus on geodesics for particles on the brane, where the brane metric is obtained through a projection $d\Omega_n^2=0$.
A classical particle stuck on the brane follows a geodesic curve $x^\mu(\lambda)$, with parameter $\lambda$, determined by varying the action
\begin{equation}\label{variational_geodesics}
S=\int d\lambda\left(\dfrac{1}{2}\dfrac{dx^{a}}{d\lambda}\dfrac{dx_{a}}{d\lambda}\right) \; .
\end{equation}
In this formulation, the conserved quantities are identified by looking at the symmetries of the Lagrangian, or equivalently, the Killing vectors of the metric. The brane metric, given by the projection of~\eqref{MP_metric}, has the same form as the Kerr metric. This type of metric produces two obvious conserved quantities associated with its time and azimuthal Killing vectors and a third one related to a Killing tensor. For example in~\cite{Creek:2007sy} the three conserved quantities were combined with the Hamiltonian (which is also conserved since the Lagrangian~\eqref{variational_geodesics} does not depend on $\lambda$) to obtain a radial equation of motion for a particle with mass $\mu$. We can apply the same reasoning to our case, where we will consider more general trajectories. For simplicity we switch to horizon radius units where $r_H=1$ so that $\bar{\mu}\rightarrow 1+a^2$ and we have the following mapping of parameters
\begin{equation} \label{eq:new_units}
\dfrac{r}{r_H}\rightarrow r \hspace{1cm} a_\star=\dfrac{a}{r_H}\rightarrow a \hspace{1cm} \omega r_H \rightarrow \omega \hspace{1cm}  \mu r_H \rightarrow \mu 
 \; ,
\end{equation}
The equations for the geodesics are then
\begin{equation}
\begin{array}{rl}
\Sigma\dfrac{dt}{d\lambda}=&\dfrac{r^2+a^2}{\Delta}\left[\left(r^2+a^2\right)\omega-a \ell_z\right]+a \ell_z-a^2\omega\sin^2{\theta} \vspace{3mm}\\
\Sigma\dfrac{d\phi}{d\lambda}=&\dfrac{a \omega\left(r^2+a^2\right)-a^2 \ell_z}{\Delta}-a\omega+\dfrac{\ell_z}{\sin^2{\theta}} \vspace{3mm}\\
\left[\Sigma\dfrac{d\theta}{d\lambda}\right]^2=&\mathcal{Q}-\cos^2\theta\left[a^2\left(\mu^2-\omega^2\right)+\dfrac{\ell_z^2}{\sin^2{\theta}}\right] \vspace{3mm}\\
\left[\Sigma\dfrac{dr}{d\lambda}\right]^2=&\left[\omega(r^2+a^2)-\ell_z a\right]^2-\Delta\left[\mu^2r^2+\left(\ell_z-a\omega\right)^2+\mathcal{Q}\right]
\end{array}
\end{equation}
where $\omega,\ell_z,\mathcal{Q}$ are the constants of motion associated with time translations, azimuthal translations and the Killing tensor respectively.
Next we redefine $\lambda\rightarrow \lambda \omega$, express the remaining constants of integration in terms of the impact parameter $b$ (in units of $r_H$), the polar angle of incidence $\vartheta$, the angular momentum magnitude $\ell$, the angular momentum orientation at infinity $\zeta$, and the reduced mass $\nu = \mu /\omega$ :
\begin{equation}\label{reduced_eq_motion}
\begin{array}{rl}
\Sigma\dfrac{dt}{d\lambda}=&\dfrac{r^2+a^2}{\Delta}\left[r^2+a^2-ab_z\right]+a b_z-a^2\sin^2{\theta} \vspace{3mm}\\
\Sigma\dfrac{d\phi}{d\lambda}=&\dfrac{a\left(r^2+a^2\right)-a^2 b_z}{\Delta}-a+\dfrac{b_z}{\sin^2{\theta}} \vspace{3mm}\\
\left[\Sigma\dfrac{d\theta}{d\lambda}\right]^2=&\mathcal{P}-\cos^2\theta\left[a^2\left(\nu^2-1\right)+\dfrac{b_z^2}{\sin^2{\theta}}\right] \vspace{3mm}\\
\left[\Sigma\dfrac{dr}{d\lambda}\right]^2=&\left[r^2+a^2-a b_z\right]^2-\Delta\left[\nu^2r^2+\left(b_z-a\right)^2+\mathcal{P}\right]\equiv \mathcal{R}
\end{array}
\end{equation}
where
\begin{equation}
\begin{array}{ccc}\label{reduced_vars}
b\equiv\dfrac{\ell}{\omega}, &b_z\equiv\dfrac{\ell_z}{\omega}=b\cos{\zeta}\sin{\vartheta}, &\mathcal{P}\equiv b^2-b_z^2-a^2\cos^2{\vartheta}
\end{array} \; .
\end{equation}
The impact parameter $b$ corresponds to the distance of closest approach to the origin, if the spacetime was flat. $\zeta$ is the polar angle of the impact parameter on the plane perpendicular to the direction of incidence (see figure~\ref{fig:diagram_zeta}) or equivalently the angle between the angular momentum of the black hole and the angular momentum of the incident particle.

The most general trajectory is parametrised by the set $\{b,\zeta,\vartheta,a,\nu\}$. Our goal is to determine whether a particle is absorbed or not for a given set. This can be achieved by looking at the radial equation of motion in~\eqref{reduced_eq_motion}, and noting that $\mathcal{R}$ must be non-negative over the trajectory. Let's start by defining the functions $A,B,C$ through
\begin{equation}
\mathcal{R}= \dfrac{1}{r^{n}}\left[-Ab^2+2Bb+C\right] \; .
\end{equation}
 To investigate whether a particle is absorbed or not, we need to know if $\mathcal{R}$ takes negative values at any point of the trajectory. If so, there is a region which is inaccessible. A way to guarantee absorption is to choose the value of $b$ in an interval such that $\mathcal{R}$ is positive over the whole space. This requires determining its zeros with respect to $b$ at all points~$r$ and finding the intervals of $b$ such that such zeros do not exist over the whole radial domain. In principle this may produce complicated discs or ring-like regions on the $\left\{b,\zeta\right\}$ polar plane at infinity. In what follows we focus on $\nu=0$, though the treatment can be easily generalised. This is because $\nu = \mu/\omega$, so since the geometrical description is good for $\omega$ large, this parameter should be small. The detailed analysis in appendix~\ref{app:Critical} leads to the following result. The range of impact parameters for absorption can be summarised as
\begin{equation}\label{def_min_impact}
0<b<b_{c}\equiv\min_{\{r\ge r_{low}\}}\left\{\dfrac{C}{\sqrt{B^2+AC}-B}\right\}
\end{equation}
with
\begin{equation}
r_{low}\equiv \left\{\begin{array}{cc}
r_A, & b_z\leq 0 \\
1, & b_z> 0 \end{array}\; , \right.
\end{equation}
and $r_A$ defined such that $\left.A\right|_{r_A}=0$.

Our goal is to obtain the absorptive discs as seen from infinity in various directions of observation. The most efficient way is to numerically determine the critical impact parameter for each set of parameters through the minimisation of~\eqref{def_min_impact}. Another useful approach is to solve the problem analytically in a particular case and expand around it perturbatively. This method is useful as a check of the numerical analysis and shows most of the features of the result. Equation~\eqref{def_min_impact} is easy to minimise when $a=0$ (the Schwarzschild case), leading to
\begin{equation}\label{a_0_sol}
\begin{array}{cc}
r_0\equiv\left(\dfrac{n+3}{2}\right)^{\frac{1}{n+1}} \;, \; \;&b_{0}\equiv\left(\dfrac{n+3}{2}\right)^{\frac{1}{n+1}}\sqrt{\dfrac{n+3}{n+1}} \; ,
\end{array}
\end{equation}
$r_0,b_0$ are the minimiser and minimum respectively. For small rotation parameters, it should be possible to find a good approximate solution of \eqref{def_min_impact}, by expanding perturbatively around \eqref{a_0_sol}, i.e. 
\begin{equation}\label{def_x_min}
r_c\equiv\sum_{p=0}^{+\infty}\frac{r_p}{p!} a^p \;, \; \; \; b_c\equiv\sum_{m=0}^\infty \frac{b_m}{m!}a^m ,
\end{equation}
with $r_c,b_c$ the minimiser and minimum for $a\neq 0$, respectively. These expansions can be inserted in~\eqref{def_min_impact} as to find the coefficients order by order in perturbation theory. The recurrence relations are provided in appendix~\ref{app:Recurrence}.

Figures~\ref{pert_check_plots1} and~\ref{pert_check_plots2} show a perfect agreement between the pertubative method and the numerical method, for small $a$ and they are good up to $a\sim 1$. For larger $a$, the perturbative expansion seems to hold as an asymptotic series. The inclusion of higher order corrections degrades the result and the closest we get from the exact numerical result is by keeping $O(2)$ corrections (see figure~\ref{pert_check_plots2}). However, even though the perturbative result fails for large $a$, it is consistently larger than the numerical one, which is supposed to be the true minimum.
\begin{figure}[t]
   \centering
   \includegraphics[scale=0.65,trim= 0cm 0 0cm 0,clip=true]{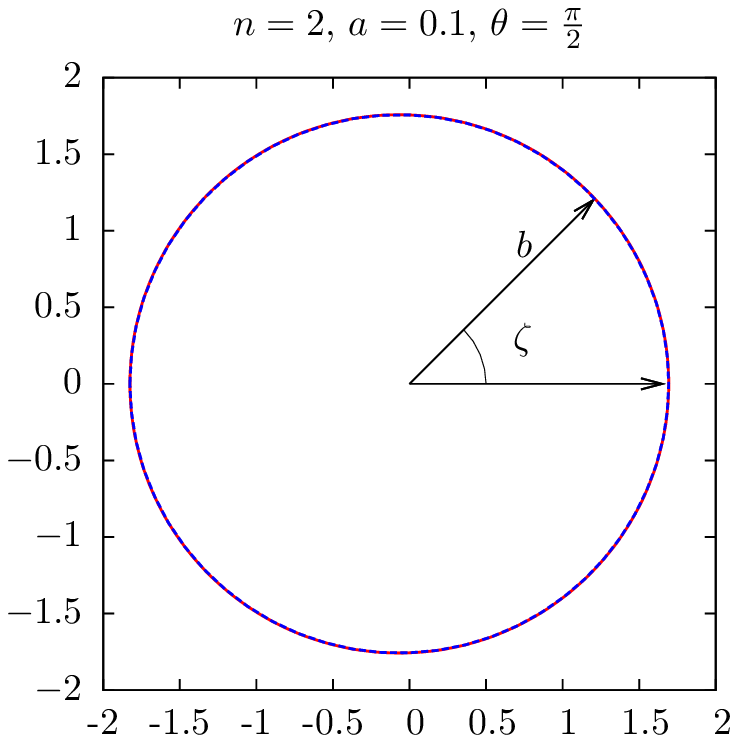}
   \includegraphics[scale=0.65,trim= 0cm 0 0cm 0,clip=true]{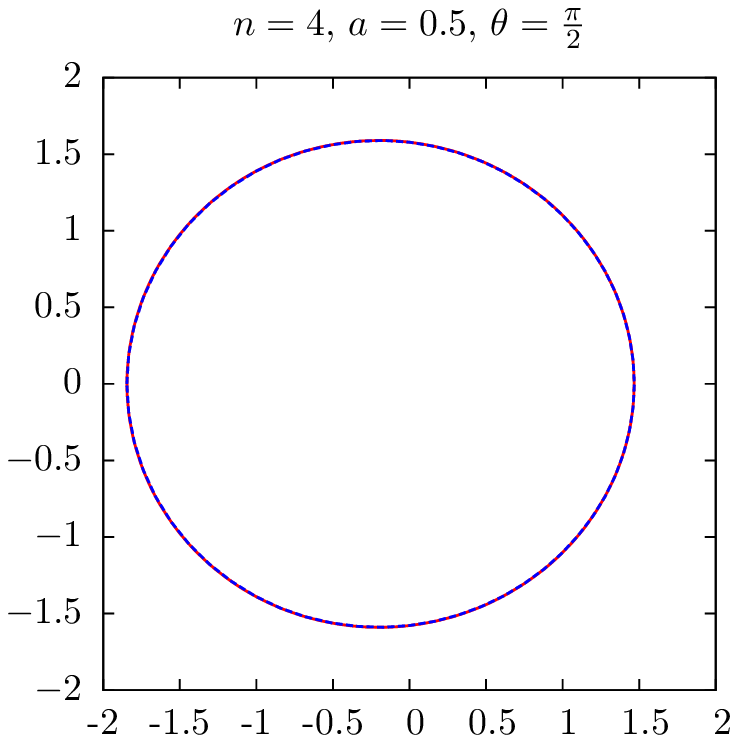}
   \includegraphics[scale=0.65,trim= 0cm 0 0cm 0,clip=true]{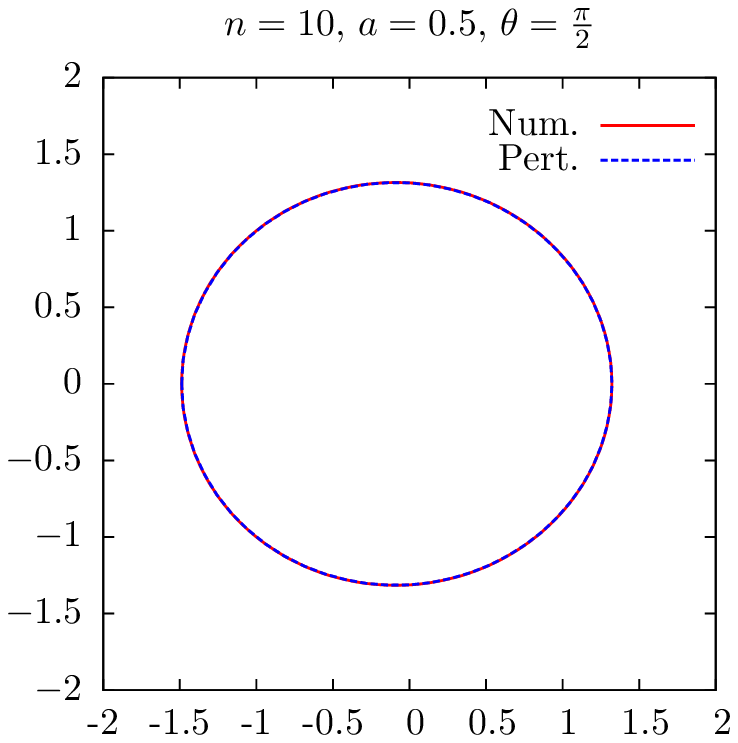}
   \caption{The perturbative result (dashed blue) agrees with the numerical minimisation (solid red), for $a\lesssim1$ (some combinations of $(n,a,\vartheta)$ indicated). The polar system $(b,\zeta)$ used throughout to draw the absorptive discs (defined on the plane transverse to the radial direction of observation at infinity) is indicated on the left plot.}
   \label{pert_check_plots1}
\end{figure}
\begin{figure}[t]
   \centering
  \includegraphics[scale=0.59,trim= 0.2cm 0 0.8cm 0,clip=true]{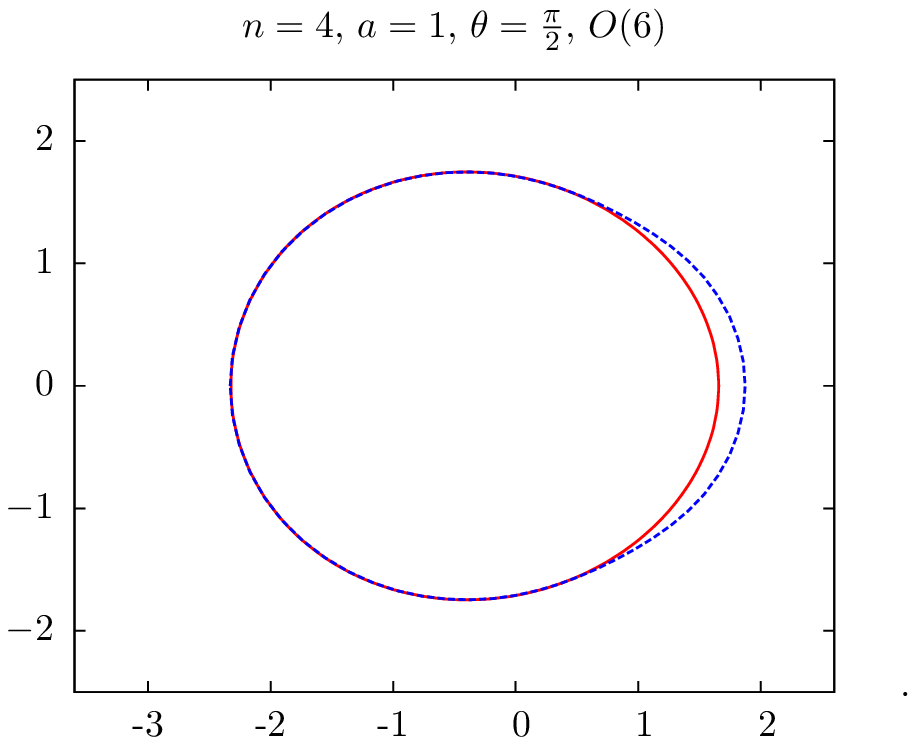}
   \includegraphics[scale=0.59,trim= 0.2cm 0 0.7cm 0,clip=true]{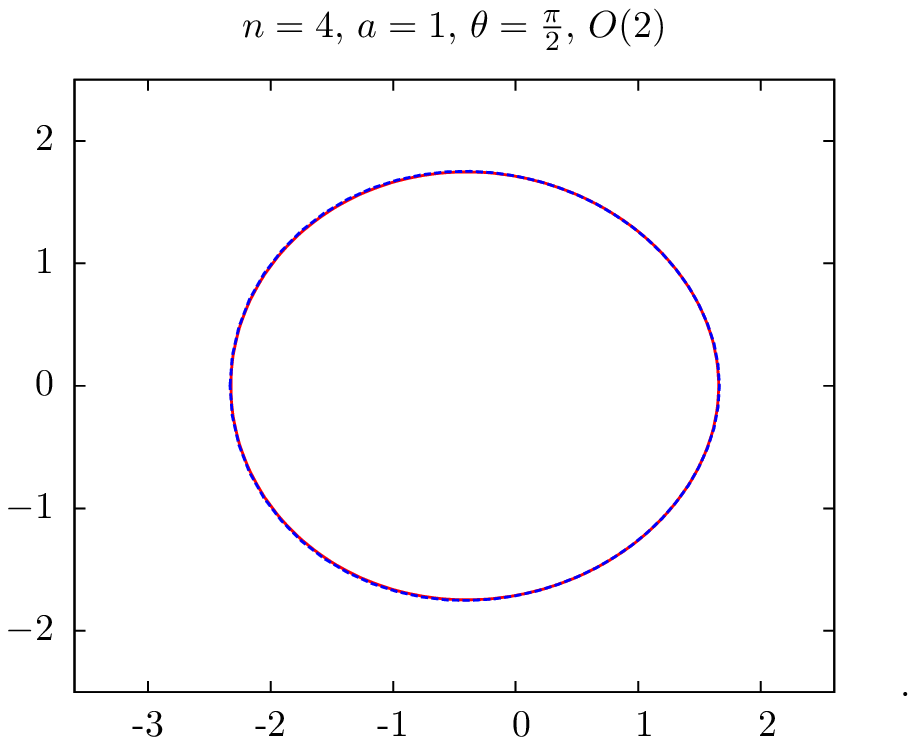}
   \includegraphics[scale=0.59,trim= 0.2cm 0 0.7cm 0,clip=true]{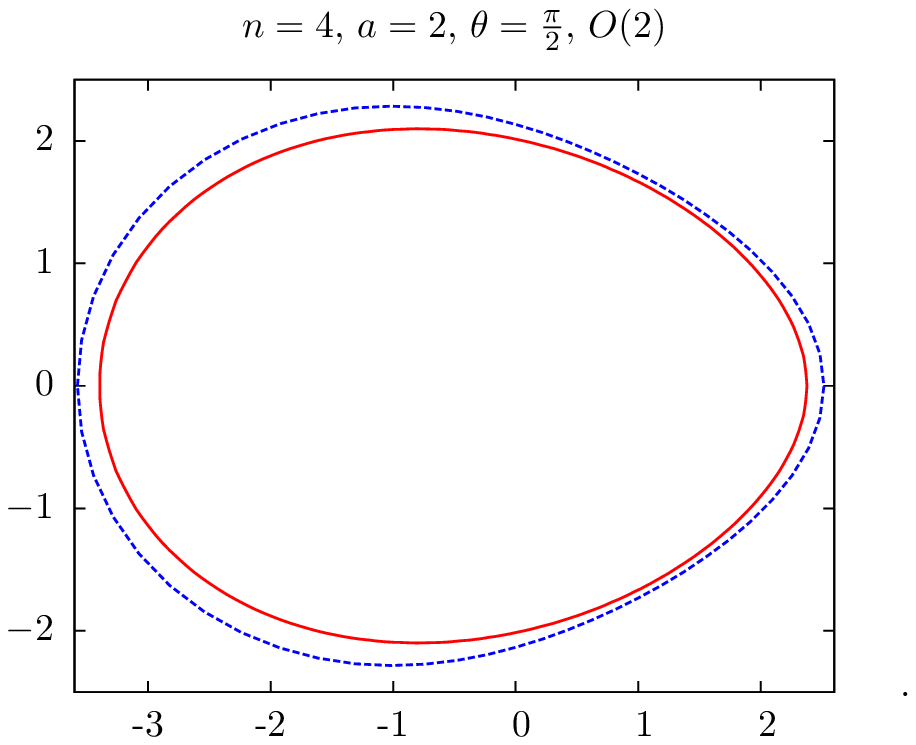}
   \caption{The perturbative result starts to fail compared to the numerical minimisation, for $a\gtrsim1$ (same color scheme as figure~\ref{pert_check_plots1}).  $n=4,\,a=1$ (\emph{left}) starts to disagree if we include $O(6)$ corrections. However, truncating at $O(2)$ (\emph{centre}) the result is still good. Increasing $a=2$  (\emph{right}) further degrades the perturbative result.}
   \label{pert_check_plots2}
\end{figure}
The lowest order correction shows that for small $a$, the distortion is suppressed for large number of extra dimensions. Furthermore, for incidence along the vertical axis, the distortion is independent of $\zeta$ (this follows trivially from the azimuthal symmetry). For incidence along the equatorial plane (or intermediate angles), the disc of absorption is distorted into an oval, indicating the orientation of the angular momentum of the black hole.
\begin{figure}
   \centering
   \includegraphics[scale=0.59,trim=0cm 0cm 0cm 0cm,clip=true]{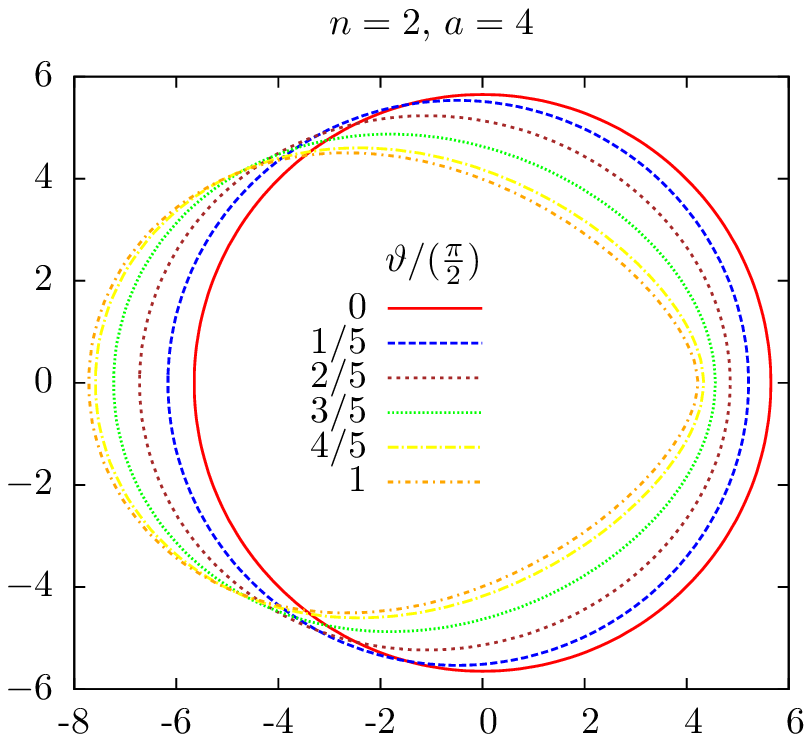} 
   \includegraphics[scale=0.59,trim=0cm 0cm 0.6cm 0cm,clip=true]{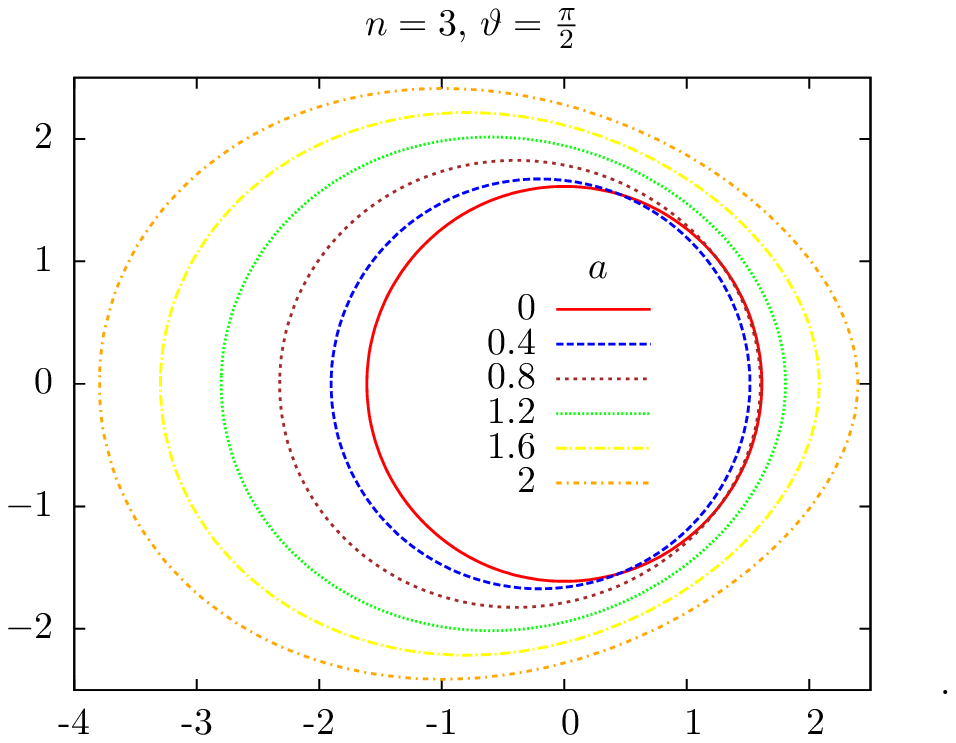}
   \includegraphics[scale=0.59,trim=0cm 0cm 0.5cm 0cm,clip=true]{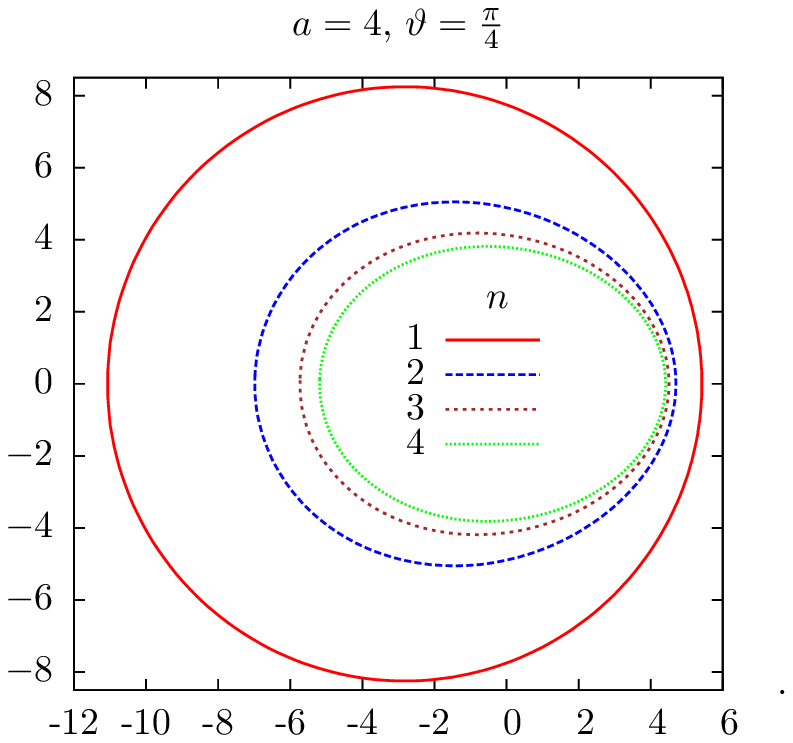} 
   \caption{The plots show the variation of the absorptive disc by varying one parameter with all the others fixed.}
   \label{xsections_plot}
\end{figure}
The exact numerical results\footnote{These were obtained by running a code in Maple which was written using the Maple inbuilt minimisation routine.} in figure~\ref{xsections_plot} confirm the non-perturbative validity of these qualitative features. 
The only exception is the conclusion regarding the suppression of the distortion with $n$ (actually the opposite occurs for $a_\star\gtrsim 1$, see right plot in \ref{xsections_plot} where the distortion is greater for larger $n$ when $a=4$). This figure also shows that for larger rotation parameter, the distortion becomes much stronger.

To better understand the contributions to the fluxes at high energy, it is instructive to look into an approximation based on the geometrical cross sections obtained from the discs above. In~\cite{Unruh:1976fm}, Unruh proved that the absorption cross section for a plane wave, incident on a Schwarzschild black hole from infinity is\footnote{We are using horizon radius units.}
\begin{equation}
\sigma=\sum_{j,m} \dfrac{\pi}{\omega^2}\mathbb{T}_{j,m}\equiv\sum_{j,m} \sigma_{j,m}.
\end{equation}
This suggests the interpretation of $\sigma_{j,m}$ as the contribution from a partial absorption cross section for a wave with angular momentum quantum numbers $\left\{j,m\right\}$.  These cross sections are directly related to the transmission factors so they are usually named greybody factors, because they are responsible for distorting the black body spectrum of Hawking radiation. In this non-rotating limit the flux becomes
\begin{equation}\label{schw_flux}
\dfrac{dN}{ dtd\omega}=\dfrac{1}{2\pi^2}\dfrac{\omega^2}{\exp(\omega/T_H)-(-1)^{2s}}\sum_{j,m}{\sigma_{j,m}}.
\end{equation}
so the Planckian term factors out of the sum.
In the limit $\omega \rightarrow +\infty$ we would expect an incident plane wave to be well described by a beam of classical particles. Then the absorption cross section is simply the area of the absorptive disc at infinity (which in the non-rotating case is a circular disc). This type of approximation was noted by DeWitt~\cite{DeWitt:1975ys}, who replaced the transmission coefficient by a theta function cutting the $j$-sum in~\eqref{schw_flux} at the maximum angular momenta allowed by the absorptive disc radius ($j=b_{max}\omega$). The success of his approximation, reinforces the interpretation of $\sigma$ as the total cross section for a classical beam of particles.

\begin{figure}
   \centering
   \hspace{1cm}\includegraphics[scale=0.8,trim= 0 0 10pt 0,clip=true]{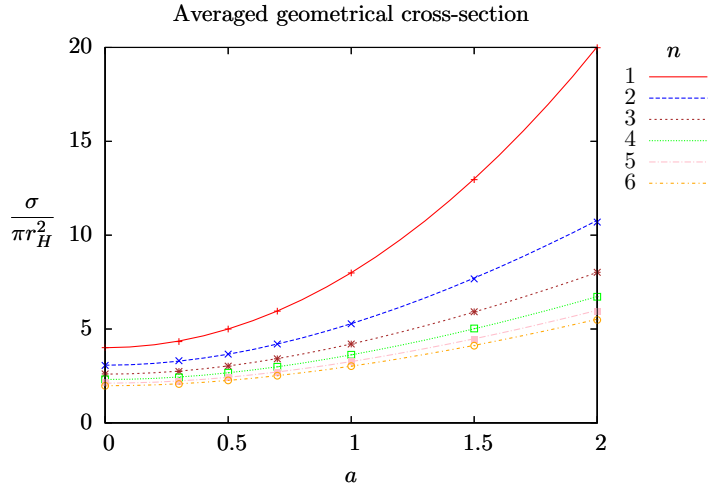}
   \caption{The plot shows a perfect agreement between the averaged geometrical cross section (solid lines) and the asymptotic value of the sum of the greybody factors (the points). We show the cases from $n=1$ (red upper line) to $n=6$ (bottom line). The curves are organise in order from the top to the bottom. We also show the $n=1$ case which could be relevant for the Randall-Sundrum model~\cite{Randall:1999vf,Randall:1999ee}, though we do not consider it in the remainder.}
   \label{plot_comparison_kanti_geom}
\end{figure}

Expression~\eqref{schw_flux} is more complicated for a rotating black hole (equation~\eqref{eq-flux-spectrum}) because the Planckian factor depends on $m$ and can not be taken out of the sum. However, it is still possible to prove that for a wave incident at an angle $\theta$, the cross section is (we are using the scalar case)~\cite{Dolan:2006thesis}
\begin{equation}
\sigma(\theta)=\dfrac{4\pi^2}{\omega^2}\sum_{k=\left\{j,m\right\}}|S_{k}(c,\cos\theta)|^2\mathbb{T}_{c,k}\equiv \sum_{k=\left\{j,m\right\}}\sigma_{c,j,m}(\theta)
\end{equation}
if we average over the solid angle we get that
\begin{equation}\label{cross_section_def}
\sigma\equiv\int{\dfrac{d\Omega}{4\pi}\sigma(\theta)}=\sum_{k=\left\{j,m\right\}}\dfrac{\pi}{\omega^2}\mathbb{T}_{c,k}= \sum_{k=\left\{j,m\right\}}\sigma_{c,j,m} \; .
\end{equation}
So for the rotating case, the same relation between partial wave cross-sections an transmission factors exists. But in the high energy limit, assuming the geometric description is good, we know how to compute the absorption cross-sections for a given angle of incidence by using the discs obtained above. Furthermore, note that at high energies we would expect this result to be independent of the spin of the particle. In figure~\ref{plot_comparison_kanti_geom}
we compare the geometrical result obtained by numerically integrating the absorptive discs, with the asymptotic value of the sum over transmission factors computed for the scalar field. We find an excellent agreement between the points taken from Table~I of~\cite{Creek:2007sy} (where a wave scattering calculation was done) and our geometrical calculation.

Furthermore, one can extend DeWitt's argument to compute the Hawking flux. If we go back to the scalar angular flux before integration over the solid angle $d\Omega$ we have
\begin{eqnarray}
\frac{d N}{dt d\omega d\Omega} &=& \frac{1}{4\pi^2} \sum_{k} \frac{1}{\exp(\tilde{\omega}/T_H) - 1} \mathbb{T}^{(4+n)}_{k}|S_{k}(c,\cos\theta)|^2 \nonumber \\
&=&\frac{\omega^2}{(2\pi)^4} \sum_{k} \frac{\sigma_{c,j,m}(\theta)}{\exp(\tilde{\omega}/T_H) - 1} \; .
\end{eqnarray}
Since we know the partial absorption cross sections as a function of the angle by using the geometrical discs, we can find a high energy approximation for the spectrum by cutting off the sums according to the allowed regions on the $\left(b,\zeta\right)$ plane.
A particularly interesting limit is when we approximate the exponential in equation~\eqref{eq-flux-spectrum} as
\begin{equation}
\tilde{\omega}\simeq \omega
\end{equation}
if
\begin{equation}
\omega \gg \left|\dfrac{a m}{1+a^2}\right|\leq\left|\dfrac{a j}{1+a^2}\right|\leq \left|\dfrac{a j_{\rm max}}{1+a^2}\right| =\left|\dfrac{a b_{\textrm{max}}\omega}{1+a^2}\right|
\end{equation}
which holds for $a$ small or large or for small enough $j$. Furthermore $j_{\mathrm{max}}$ is large, when $\omega\rightarrow+\infty$, so for most of the modes contributing to the the sum in~\eqref{eq-flux-spectrum} this approximation should work. This implies a similar factorisation in the high energy limit for rotating black holes:
\begin{equation}\label{DeWittapprox}
\frac{d N}{dt d\omega} \approx \frac{1}{2\pi^2}\frac{\omega^2}{\exp(\omega/T_H) - 1}\sigma \; .
\end{equation}

The approximations obtained in this section provide some interesting information about the Hawking spectrum in the high energy limit (where it should be independent of the spin of the particle). As expected, the discs for incidence along the angular momentum axis are circular, in agreement with the symmetry of the background. The angular asymmetry as we vary $\theta$ becomes manifest in the shape  of the discs. As verified in~\eqref{cross_section_def} and~\eqref{DeWittapprox} there is a direct correspondence between the absorptive discs and the Hawking spectrum at high energies. Another interesting property that we can see in the middle plot of figure~\ref{xsections_plot} is an enhancement of the Hawking emission in the high energy limit, as we increase the rotation parameter of the background (another known generic property of the Hawking spectrum), since the discs grow in size. Some of these properties will appear in the next section.

\subsection{Exact Hawking fluxes}

The angular spheroidal functions ${}_{s}S\left(a\omega,\cos\theta\right)$ are the main ingredient inducing non-uniform angular distributions which we shall investigate. They are obtained by solving the spheroidal wave equation as detailed in previous studies~\cite{Ida:2002ez,Harris:2005jx,Ida:2005ax,Duffy:2005ns,Casals:2005sa,Ida:2006tf,Casals:2006xp,Casals:2008pq,Casals:2009st}. To quickly recall the relevant properties for our analysis, we will simply use the data available in~\cite{greybody_data} for the transmission factors and the method in~\cite{Frost:2009cf} to evaluate the spheroidal functions and obtain the angular fluxes for different brane degrees of freedom. 

\begin{figure}
  \centering
   \hspace{-2mm}\includegraphics[scale=0.55]{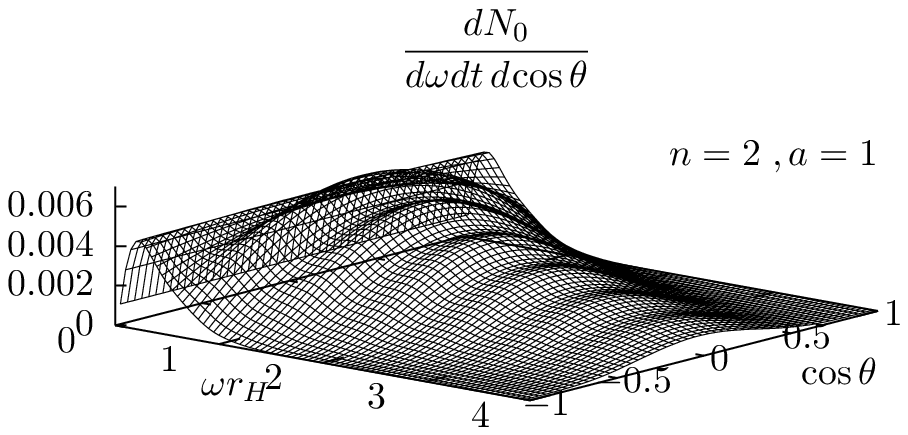} 
   \includegraphics[scale=0.55]{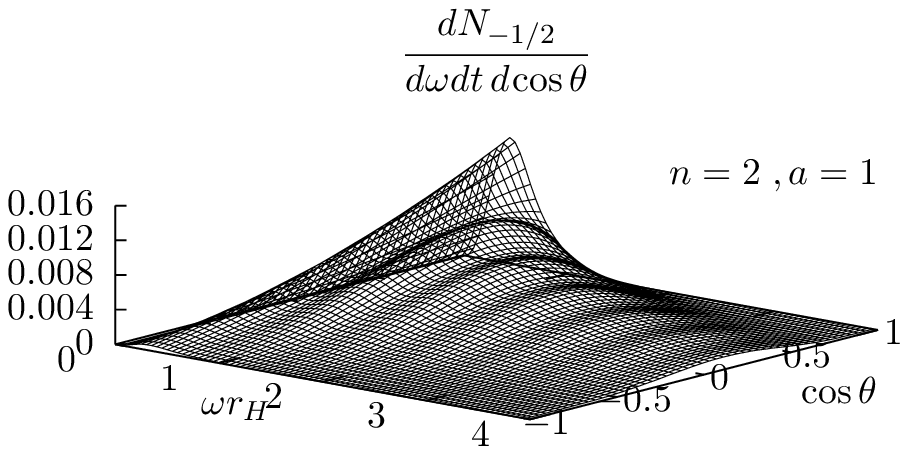} 
   \includegraphics[scale=0.55]{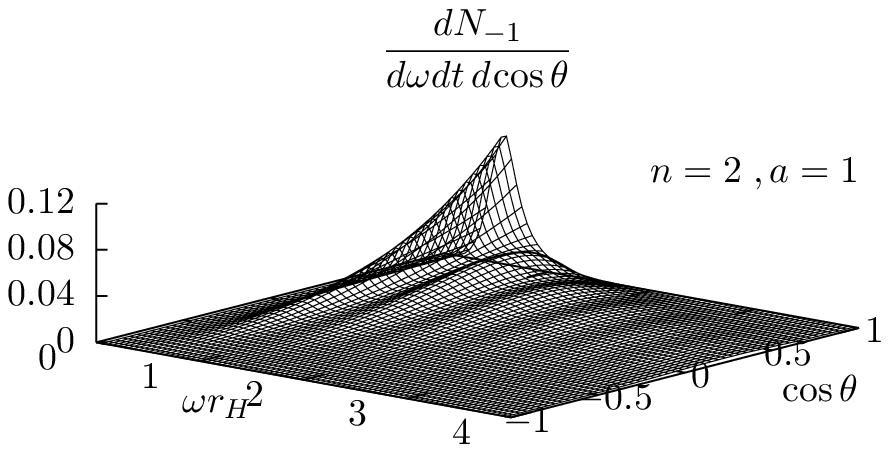} \vspace{1mm}\\
   \hspace{-2mm}\includegraphics[scale=0.55,clip=true,trim=0cm 0cm 0cm 0.9cm]{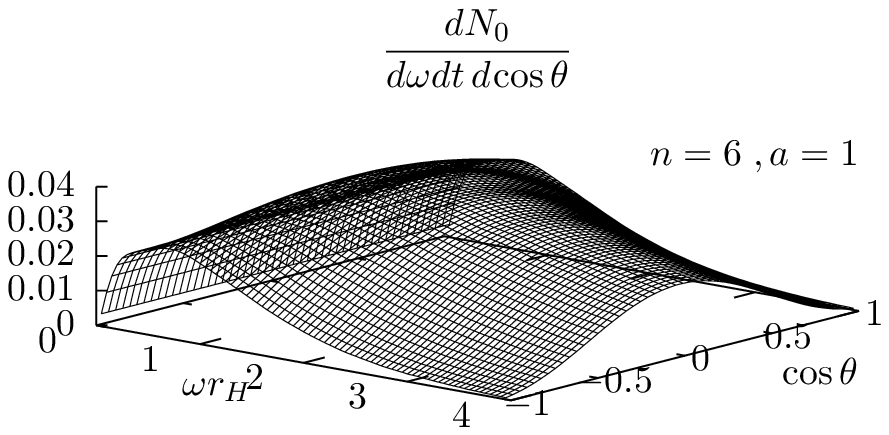} 
   \includegraphics[scale=0.55,clip=true,trim=0cm 0cm 0cm 0.9cm]{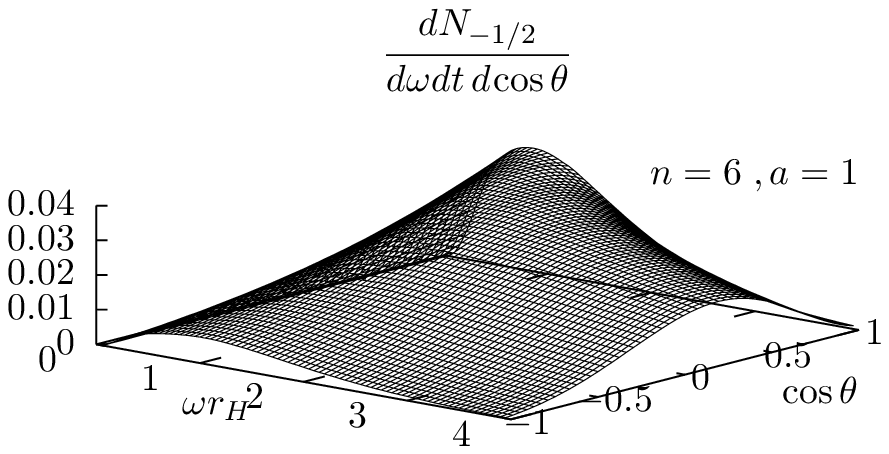} 
   \includegraphics[scale=0.55,clip=true,trim=0cm 0cm 0cm 0.9cm]{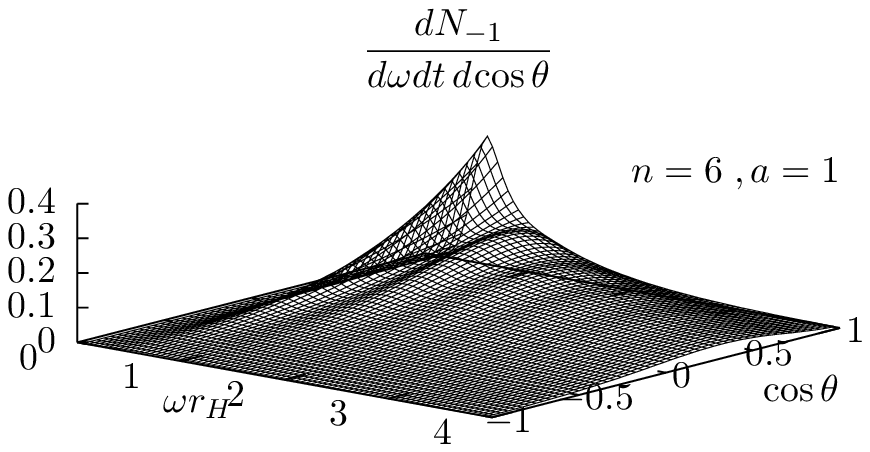} 
   \caption{\label{dN_dxdOmga} Angular dependence of the flux for states of positive helicity $h=-s=0,1/2,1$, non-zero rotation $a=1$ and two different $n=2,6$. The negative helicity plots are obtained by the reflection $\cos\theta\rightarrow -\cos\theta$} 
\end{figure}

Some important properties of the angular distributions follow from the observation of the differential number fluxes in figure~\ref{dN_dxdOmga}. 
It is well known that in general rotation tends to make the spheroidal functions more axial whereas the combination of the transmission factor and Planckian factor favour $m=j$ modes at large rotation to spin down the black hole. The competition between the angular functions and the latter results in more axial angular distributions at low energies and more equatorial distributions at high energies when rotation is on. This is seen from the energy dependence of the angular profiles shown in figure~\ref{dN_dxdOmga}. 

These effects are particularly large for vector bosons, which are more likely to be emitted close to the rotation axis at low energies, whereas high energy vector bosons are more likely to be emitted in the equatorial plane (the same happens for fermions). In figure~\ref{dN_dxdOmga} we see that particles with a single helicity will be emitted asymmetrically by a rotating black hole~\cite{Leahy:1979xi}. For example, if the black hole angular momentum vector is pointing north, then positive(negative) helicity states will be preferentially emitted in the northern (southern) hemisphere. As suggested before in~\cite{Frost:2009cf}, this might lead to angular asymmetries in the decays of unstable $W$ and $Z$ vector bosons which can be in two possible transverse polarisations states.

\section{A parton level study of angular correlations}\label{partonStudy}
The remainder of this article is dedicated to studying the feasibility of observing these angular asymmetries in a more realistic model for the decay of TeV gravity black holes at colliders. We use the recently upgraded \texttt{CHARYBDIS2} event generator which includes in full the effects of rotation for brane degrees of freedom, including the helicity effects described in the previous section. We have produced some samples to illustrate the angular effects in question. These effects should be fairly independent of the details of the sample, given that rotation is expected to be present in general, and the qualitative features of the angular fluxes described in the previous section are fairly independent of $n$.

\subsection{Angular momentum reconstruction}
\label{sec:J_rec}
The problem of determining the angular momentum axis and/or magnitude is much more difficult than reconstructing the mass of a rotating black hole. This is because the angular momentum axis, which controls the angular distributions, evolves during the evaporation, changing direction and magnitude. Furthermore, all the decay products in the laboratory frame are in general boosted with respect to the black hole centre of mass frame, and the black hole recoils between each emission. Nevertheless, we may hope to see some of the effects by boosting each event back to the partonic centre of mass frame. In principle we can determine the latter reliably if there is little missing energy emitted in the evaporation. The only missing energy in the SM comes from neutrinos and they represent a small part of the total number of degrees of freedom. However missing energy from neutrinos can arise from the secondary decay of SM heavy particles. Nevertheless the fraction of missing energy from the evaporation should not be so large (for most of the events) as to degrade the results much more compared to other factors such as the recoil. Note however that including gravitons in the evaporation may degrade the reconstruction further, though in principle, by measuring all the proton remnant pieces we may be able to obtain a good measure of the partonic centre of mass frame. Alternatively to reduce this effect we may cut on events with a small amount of missing energy as it has been done in previous analyses for the mass reconstruction (see~\cite{Frost:2009cf} and references therein).

We start our study by considering some distributions which are not accessible observables at a collider experiment, but help to understand the angular correlations involved. The aim is to understand how the momenta of the particles emitted correlate with the true angular momentum axis by using knowledge of the black hole decay history.

\begin{figure}[t]
\includegraphics[scale=0.59,clip=true,trim= 22 0 0 0]{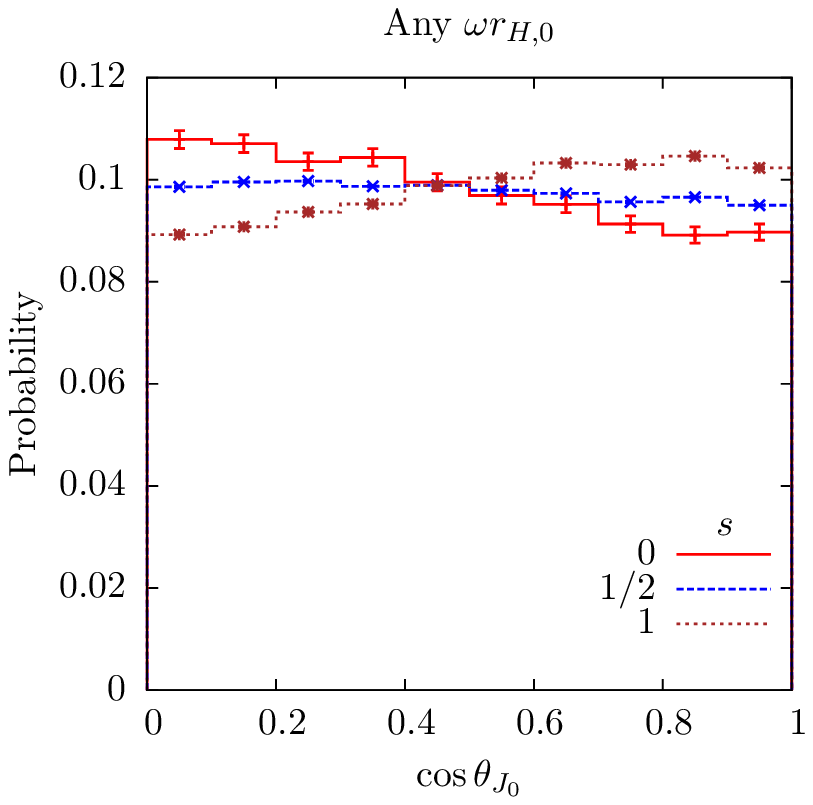} \hspace{0.5mm} 
\includegraphics[scale=0.59,clip=true,trim= 22 0 0 0]{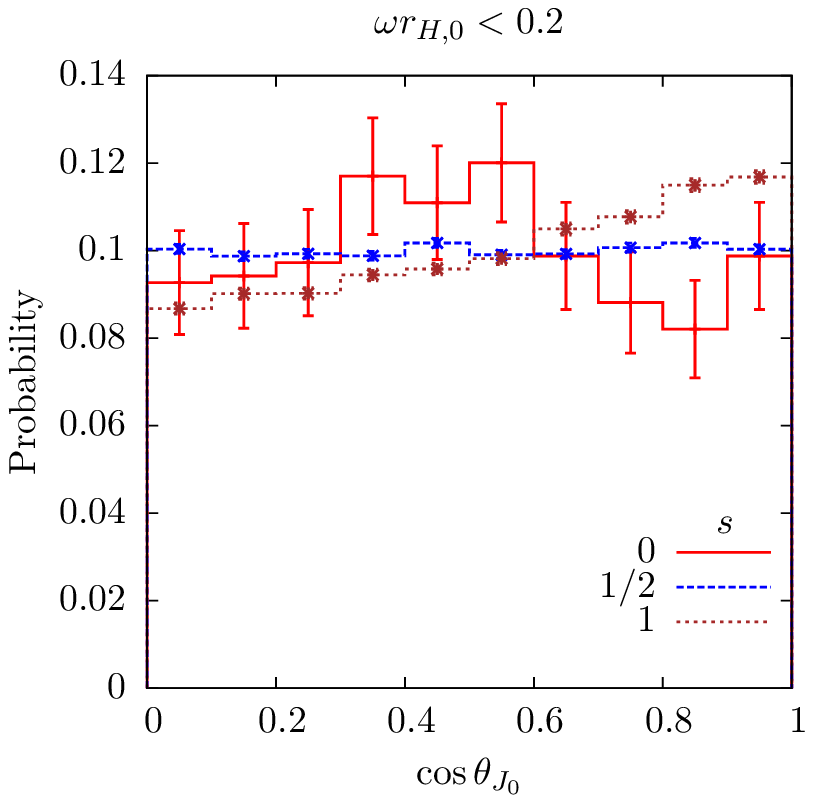} \hspace{0.5mm} 
\includegraphics[scale=0.59,clip=true,trim= 22 0 0 0]{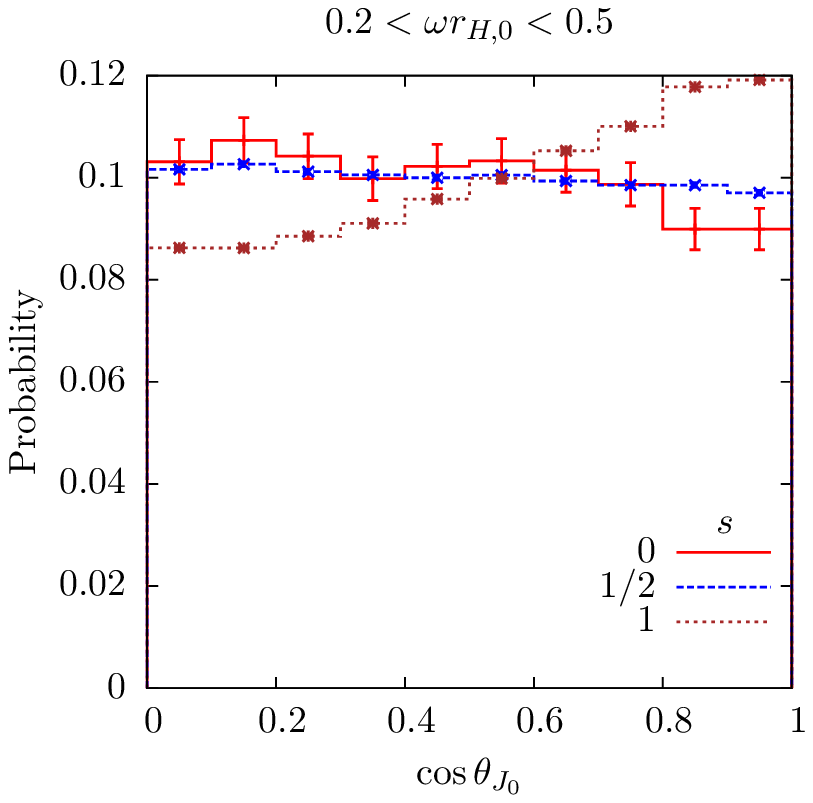} \vspace{0mm}\\
\includegraphics[scale=0.59,clip=true,trim= 22 0 0 0]{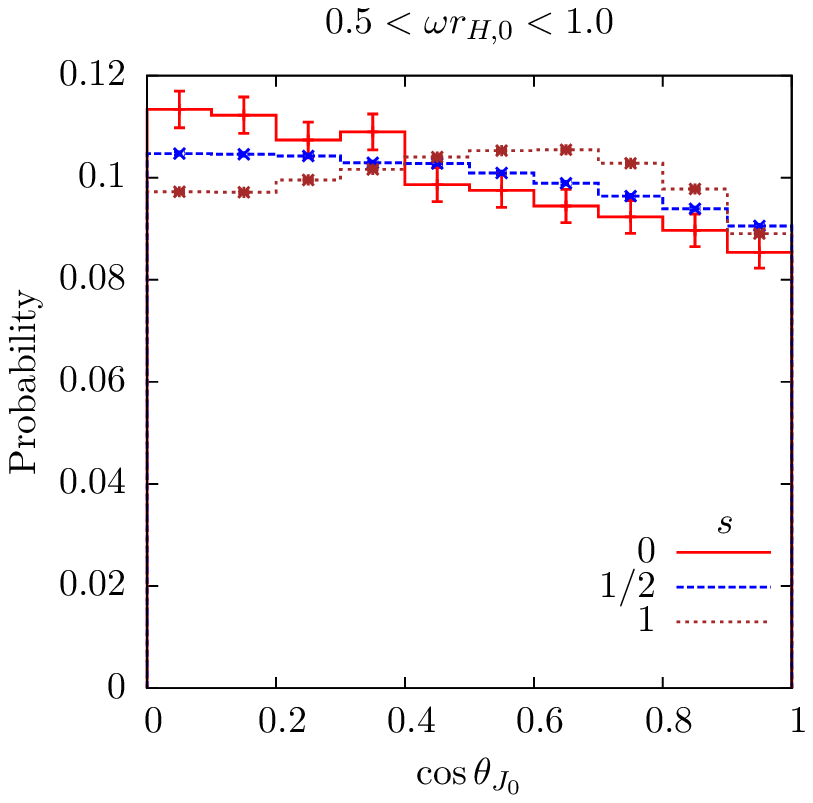} \hspace{0.5mm} 
\includegraphics[scale=0.59,clip=true,trim= 22 0 0 0]{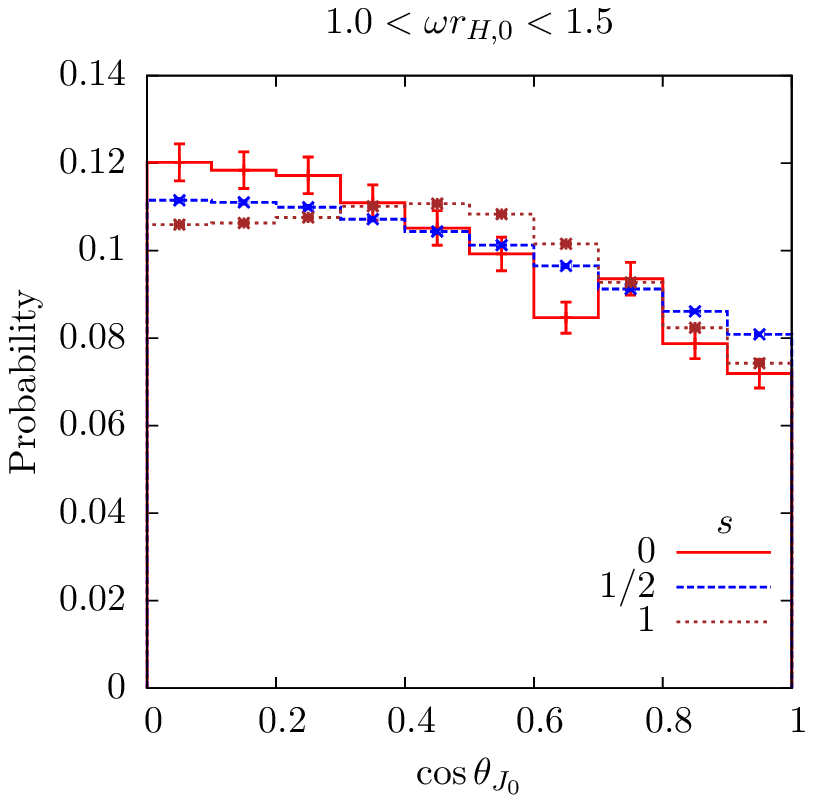} \hspace{0.5mm} 
\includegraphics[scale=0.59,clip=true,trim= 22 0 0 0]{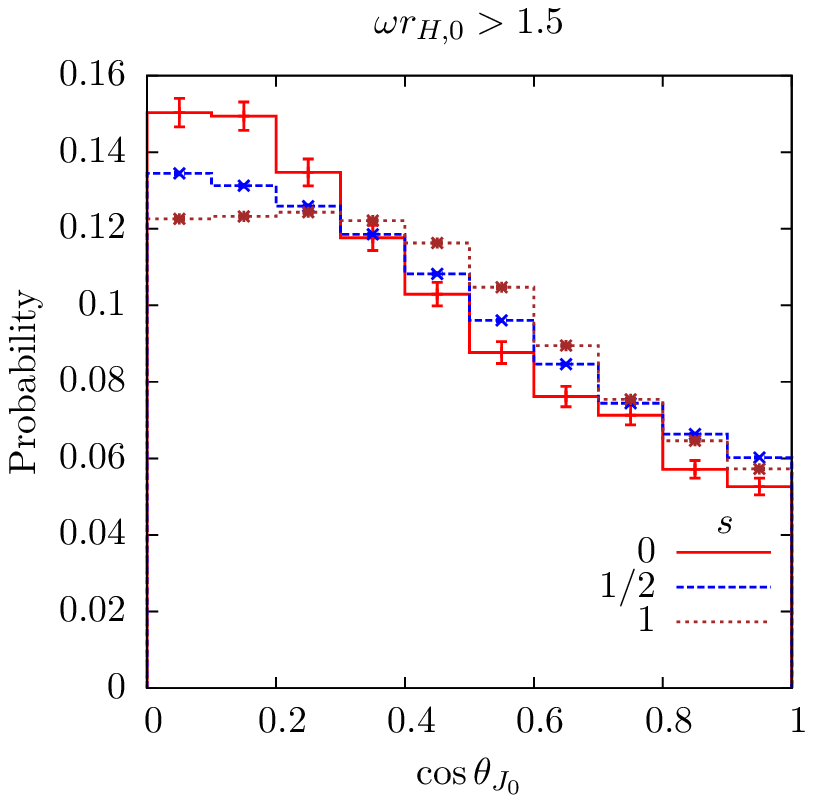}
\caption{\label{fig:Jrec_spins} {\em Distribution of angles of particles with various spins with the angular momentum axis.} The top left plot shows the distributions for $\cos \theta$ for various spins for an $n=3$ sample with black hole  masses above $5$~TeV. The other plots show the distribution for various ranges of energies as indicated in the title of each plot. Note that the error bars for $s=0$ are due to the small number of pure scalar degrees of freedom in the SM (only the Higgs particle).}
\end{figure}

The first distribution we study is the angle of a particle with a given spin with the true initial angular momentum in the centre of mass frame of the initial black hole
\begin{equation}
\cos\theta_{J_0}\equiv \dfrac{\mathbf{p}_s\cdot \mathbf{J}_0}{|\mathbf{p}_s||\mathbf{J}_0|} \; .
\end{equation} 
The first plot in figure~\ref{fig:Jrec_spins} shows a spin dependent behaviour as expected. The sample used to produced the plots used all the defaults values except for \texttt{MJLOST=.FALSE.}. This produces black hole events with larger angular momentum and mass. This is not essential since we could use a cut on the visible invariant mass for the event to select heavier black holes if \texttt{MJLOST=.TRUE.}, it is simple advantageous to generate samples faster.
We know from the theoretical plots in figure~\ref{dN_dxdOmga} that scalars and fermions tend to be more equatorial (though fermions at low energies also have a small axial peak) and vector bosons are very axial at low energies. This is consistent with the larger probability for vector bosons at larger $|\cos\theta|$ in the top left plot (though the effect is not very large). We can improve this correlation by selecting particles in particular ranges of energy for each initial black hole. For example if our cut requires high energy particles for all spins we would expect an equatorial correlation, whereas at low energies we expect a flatter distribution for scalars and fermions and an axial distribution for vector particles. This is confirmed in the two top right plots and the bottom plots of figure~\ref{fig:Jrec_spins} where we have chosen ranges of energy in units of the horizon radius of the initial black hole. This result suggests using soft vector particles as a guess for the axis, to plot angular distributions. This will fail a considerable part of the time and smear out the true correlation. 

\begin{figure}[t]
\centering
\includegraphics[scale=0.63,clip=true,trim= 22 0 0 0]{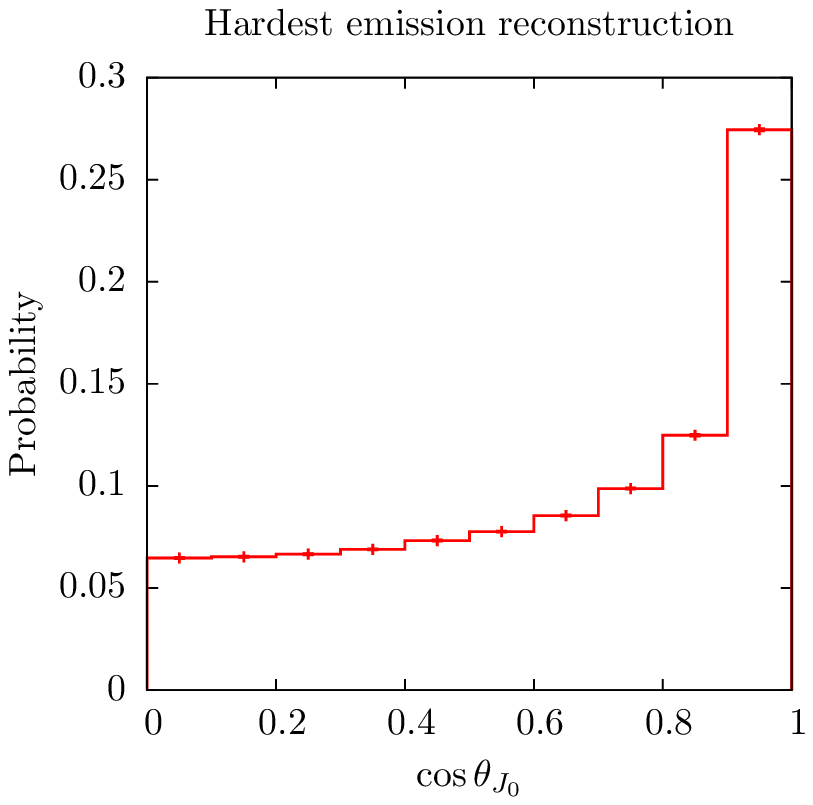} \hspace{2cm} 
\includegraphics[scale=0.63,clip=true,trim= 22 0 0 0]{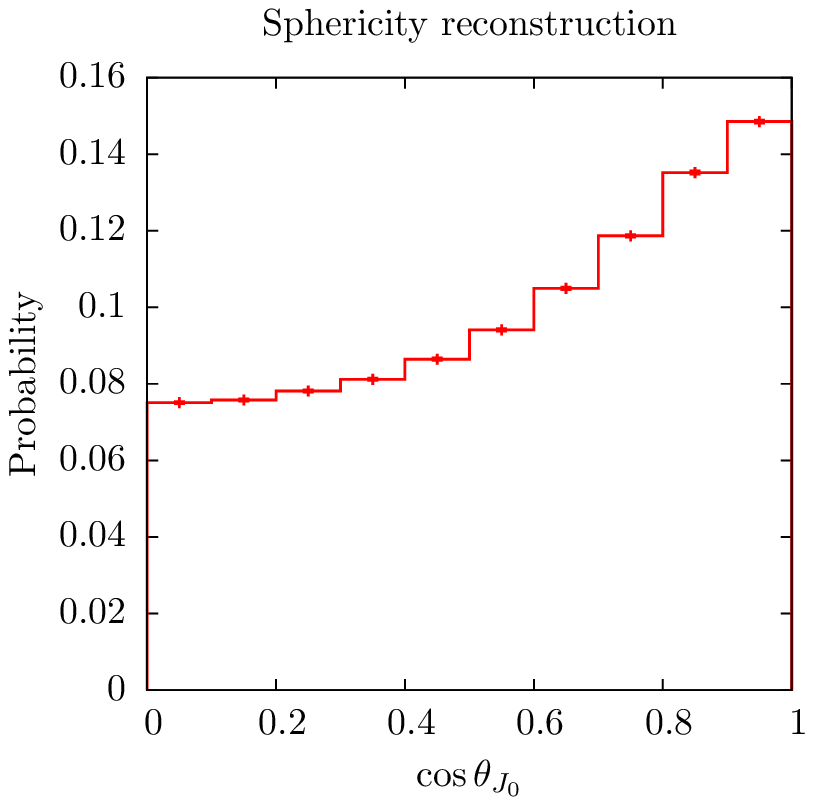} 
\caption{\label{fig:Jrec_perp_shapes} {\em Axis reconstruction:} Using the hardest emission (left) and the eigenvector of the smallest eigenvalue of the sphericity tensor (right). }
\end{figure}

High energy particles tend to be emitted perpendicularly to the initial black hole angular momentum. This is seen in the bottom right plot of figure~\ref{fig:Jrec_spins} where the correlation is stronger.  If in addition we assume that the direction of the angular momentum vector is perpendicular to the direction of the black hole momentum (which is true in the limit where the angular momentum does not recoil during the production) we obtain another guess for the axis. The left plot in figure~\ref{fig:Jrec_perp_shapes} shows that this method works better. However it relies on the assumption that the initial angular momentum is perpendicular to the black hole momentum.

A third method to estimate the angular momentum axis is to consider the shape of the event, i.e. to use all the momenta in the decay. For a rotating black hole we expect most of the particles to be emitted equatorially (since only low energy vector bosons are axial). Thus the event should have a disc like distribution of momenta indicating the orientation of the axis. This axis should minimise the amount of momentum projected along its direction. Another advantage of this reasoning is that it gives lower weight to low energy particles which we want to eliminate since low energy vector bosons are more axial, and low energy scalars and fermions are more uniform. If we denote the direction of the angular momentum by $\mathbf{n}$, and use projections of momenta squared, then we want to minimise
\begin{eqnarray}
\dfrac{\min_{\mathbf{n}}\sum_{i}(\mathbf{p}_{i}\cdot\mathbf{n})^{2}}{\sum_{i}\left|\mathbf{p}_{i}\right|^{2}}&=&\dfrac{\min_{\mathbf{n}}\sum_{i}\sum_{\alpha\beta}\mathbf{n}^{\alpha}\mathbf{p}_{i}^{\alpha}\mathbf{p}_{i}^{\beta}\mathbf{n}^{\beta}}{\sum_{i}\left|\mathbf{p}_{i}\right|^{2}}\nonumber\\ &=&\min_{\mathbf{n}}\sum_{\alpha\beta}\mathbf{n}^{\alpha}\left(\dfrac{\sum_{i}\mathbf{p}_{i}^{\alpha}\mathbf{p}_{i}^{\beta}}{\sum_{i}\left|\mathbf{p}_{i}\right|^{2}}\right)\mathbf{n}^{\beta} \nonumber\\&=&\min_{\mathbf{n}}\sum_{\alpha\beta}\mathbf{n}^{\alpha}S^{\alpha\beta}\mathbf{n}^{\beta}\label{eq:min_sphericity}
\end{eqnarray}
where we are use Greek letters for spatial indices and the index $i$ runs over all particles in the event. We have defined the sphericity tensor $S^{\alpha\beta}$ as usual~\cite{Bjorken:1969wi}. The sphericity tensor has the properties that all eigenvalues are non-negative and their sum is one. In the eigenbasis it is clear that the direction which minimises the quantity in~\eqref{eq:min_sphericity} is the eigenvector associated with the smallest eigenvalue.
In the right plot of figure~\ref{fig:Jrec_perp_shapes} the distribution for the angle between the guessed axis (the eigenvector associated with the smallest eigenvalue) and the angular momentum axis is shown. This method is not as good as the the left plot but it has the advantage of not relying on the assumption that the angular momentum is on the plane transverse to the collision axis.

\subsection{Angular correlators}
\label{sec:angl_corr}
\begin{figure}[t]
  \includegraphics[scale=0.68]{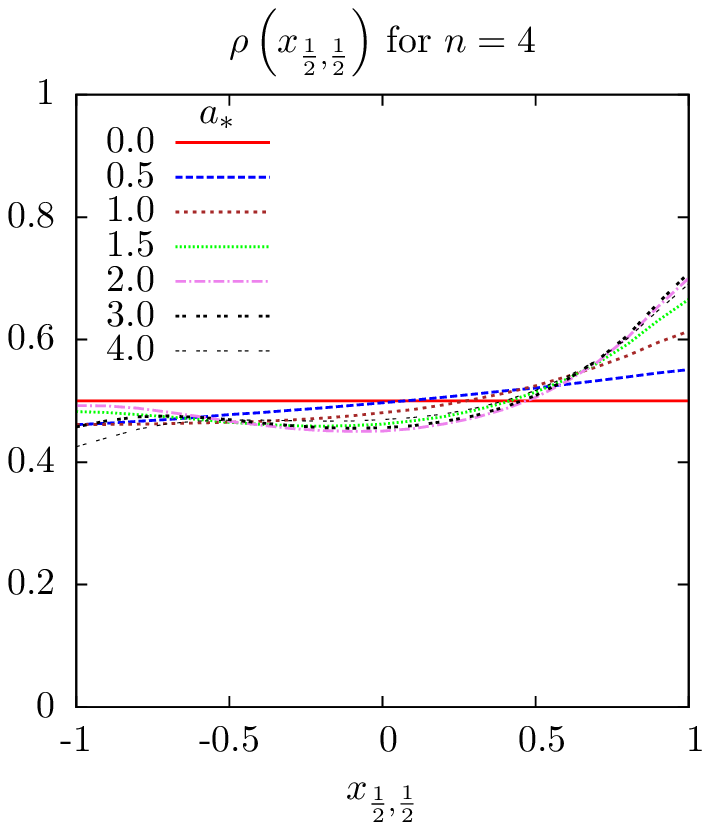} \hspace{0.5mm} 
  \includegraphics[scale=0.68]{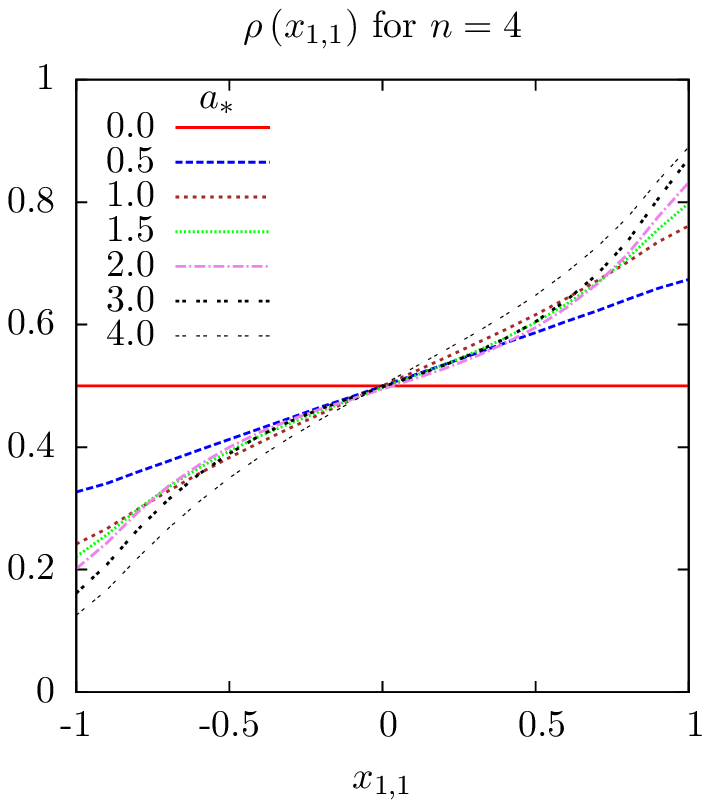} \hspace{0.5mm} 
  \includegraphics[scale=0.68]{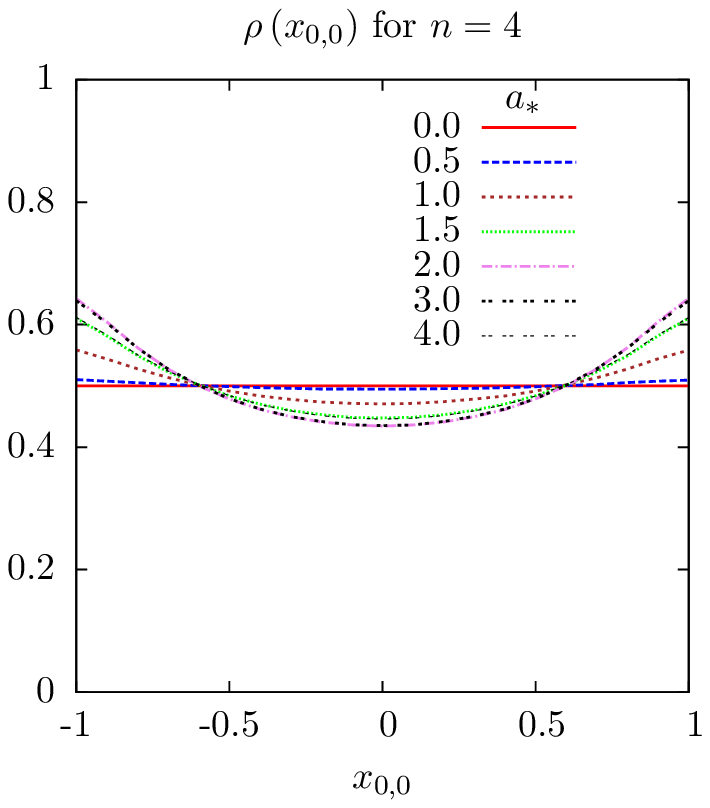} \\
  \includegraphics[scale=0.68]{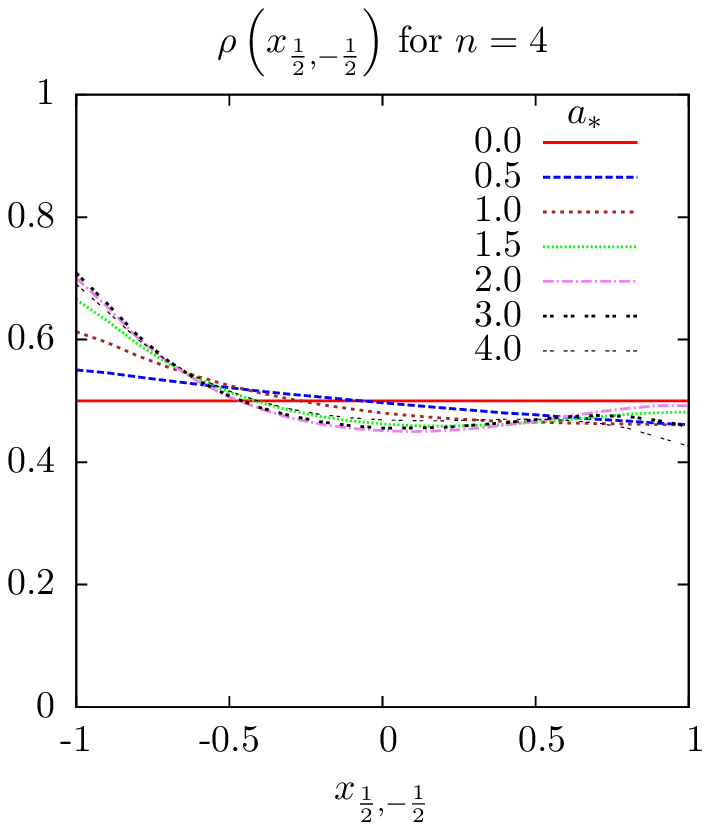} \hspace{0.5mm} 
  \includegraphics[scale=0.68]{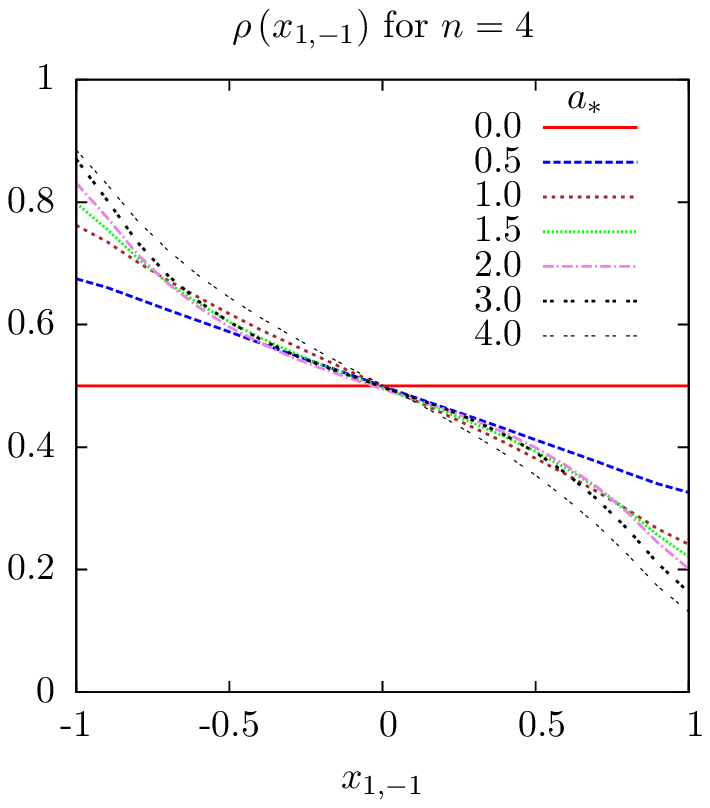} \hspace{0.5mm} 
  \includegraphics[scale=0.68]{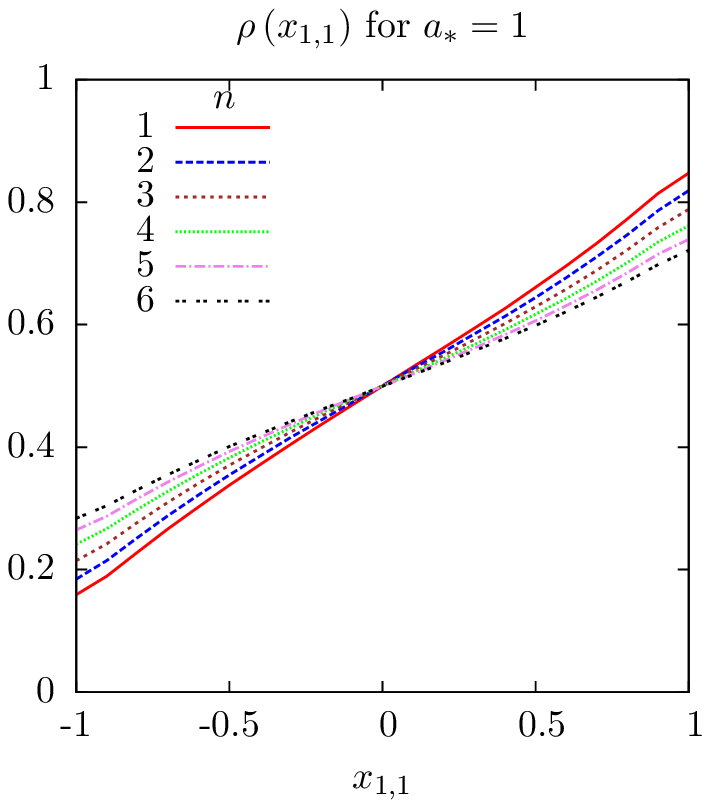}
\caption{\label{fig:th_correlators} {\em Angular correlators for an eternal black hole.} The top plots show the probability density functions for the angular correlator between particles of same helicity for fermions, vector bosons and scalars (left to right), for a range of $a_*$ values. The two left bottom plots are similar but between particles with opposite helicities. The bottom right plot shows the dependence with $n$ for $a_*=1$ and two vector bosons with the same helicity.}
\end{figure}
An alternative to reconstructing the black hole angular momentum to study angular distributions, is to explore the property that polarised angular distributions are strongly dependent on the helicity of the particle. For example from figure~\ref{dN_dxdOmga} we know that there is a strong preference for vector bosons to be emitted in different hemispheres. This motivates defining angular correlators of the form (in the frame of the initial black hole)
\begin{equation}\label{eq:x_def}
x_{i,j}=\dfrac{\mathbf{p}_i\cdot\mathbf{p}_j}{|\mathbf{p}_i||\mathbf{p}_j|}
\end{equation}
which are cosines of angles between particles $i$ and $j$. So for the case of particles with the same helicity, we would expect the distribution to be higher at $x_{i,j}\sim 1$ and reduced at $x_{i,j}\sim -1$, and the opposite to happen for particles with opposite helicities. Figure~\ref{fig:th_correlators} shows the expected behaviour for pairs of emissions in a fixed black hole background (no recoil). The probability density function used to determine the distribution of~\eqref{eq:x_def} is derived in appendix~\ref{sec:ditribs_eternal}. The effect at fixed black hole parameters grows quickly with $a_*$, especially for particles of helicity $h=\pm 1$. The bottom right plot shows that there is some variation with $n$ though not very strong (the curves are qualitatively similar), as claimed in the beginning of this section.

For an evolving black hole, we expect these effects to get smeared, due to the momentum and angular momentum recoil. Again the best procedure is to compute similar quantities in the rest frame of the initial black hole assuming a small amount of missing energy during the evaporation.

\begin{figure}
\centering
  \includegraphics[clip=true,trim=23 0 2 0,scale=0.6]{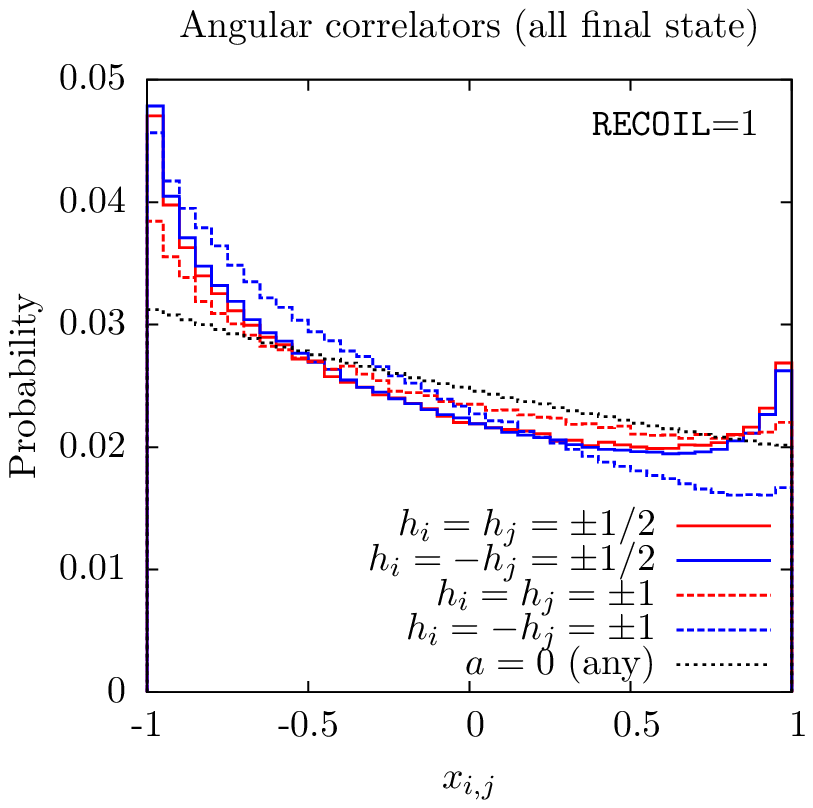}
  \includegraphics[clip=true,trim=20 0 2 0,scale=0.6]{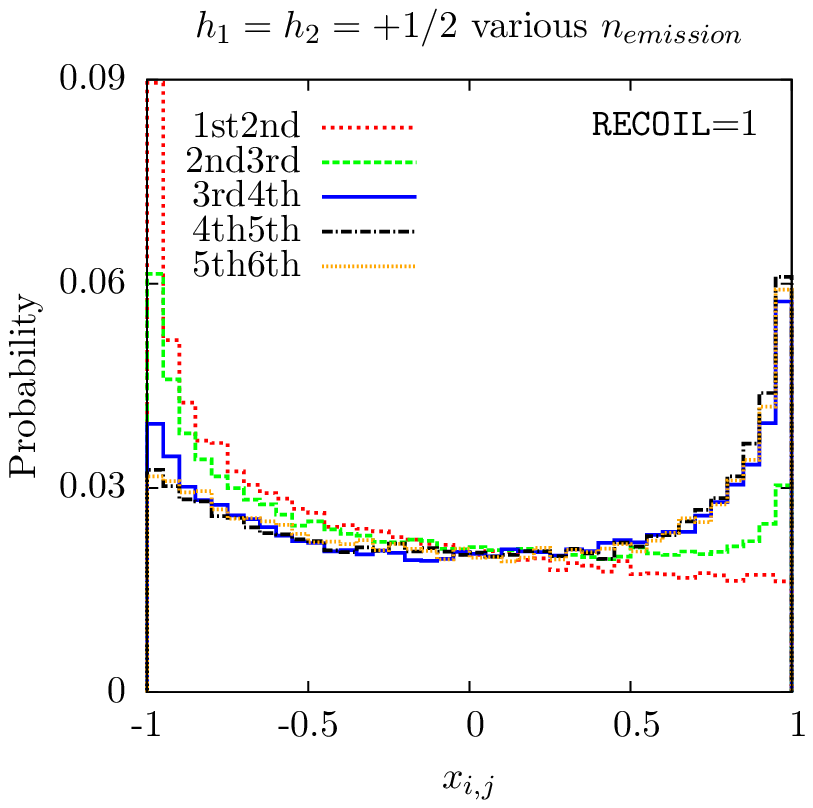}
  \includegraphics[clip=true,trim=20 0 2 0,scale=0.6]{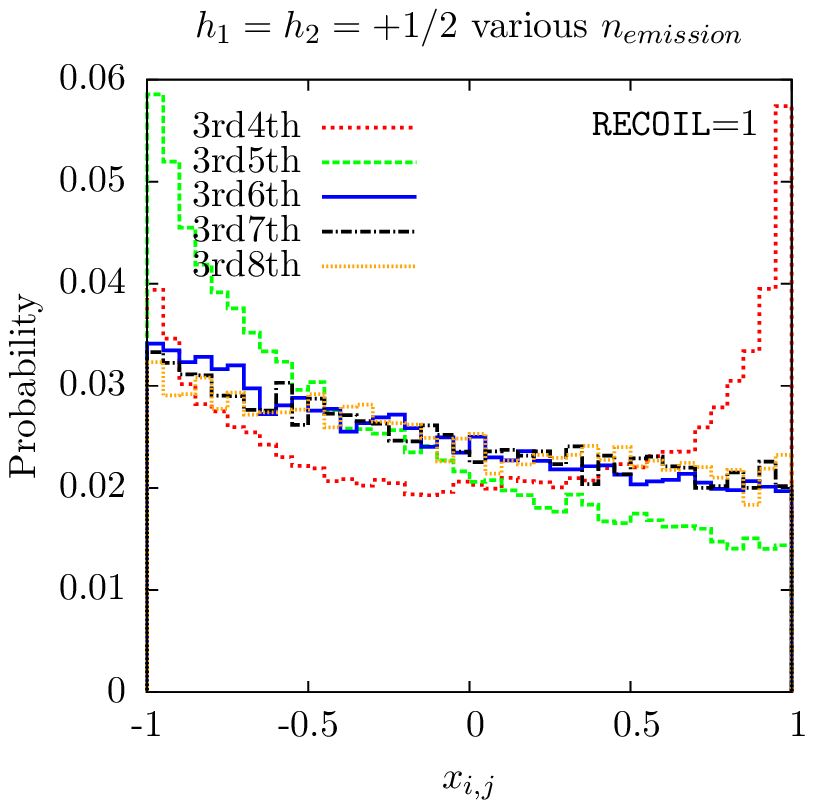} \vspace{1mm}\\
  \includegraphics[clip=true,trim=23 0 2 0,scale=0.6]{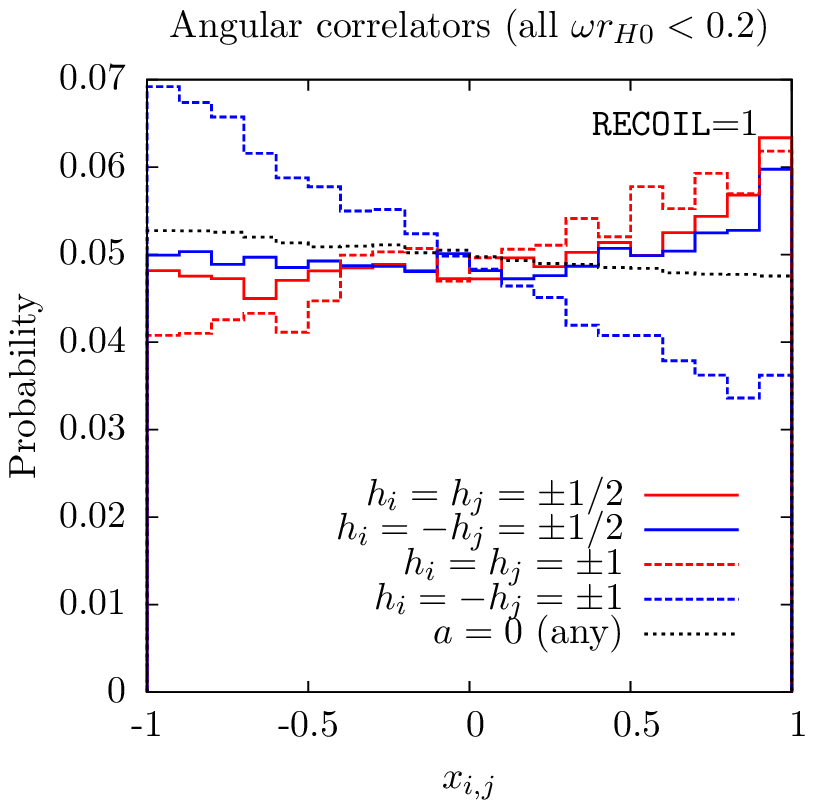}
  \includegraphics[clip=true,trim=20 0 2 0,scale=0.6]{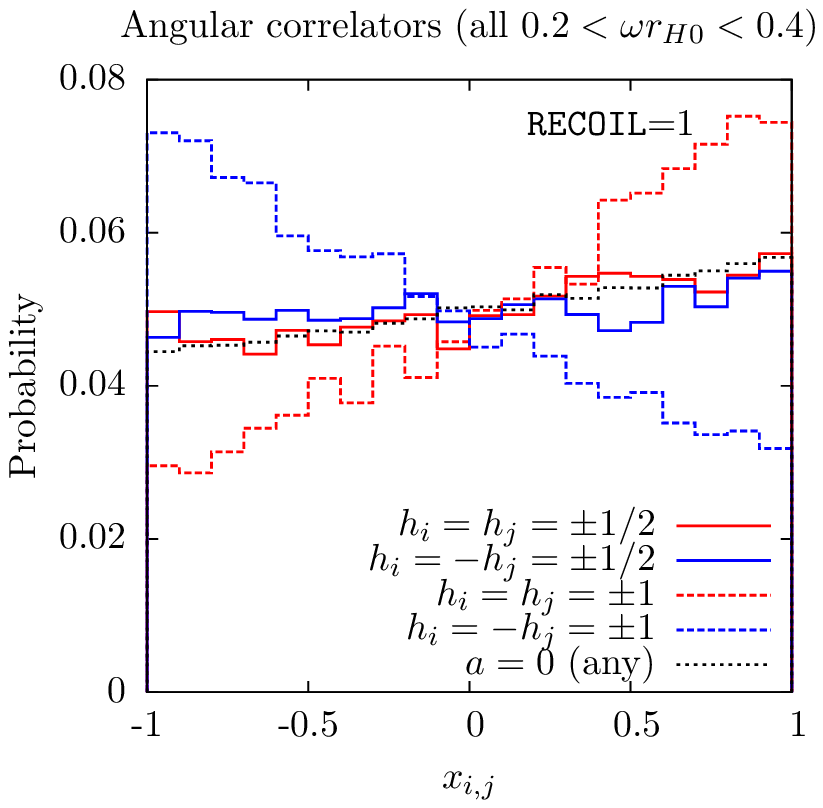}
  \includegraphics[clip=true,trim=20 0 2 0,scale=0.6]{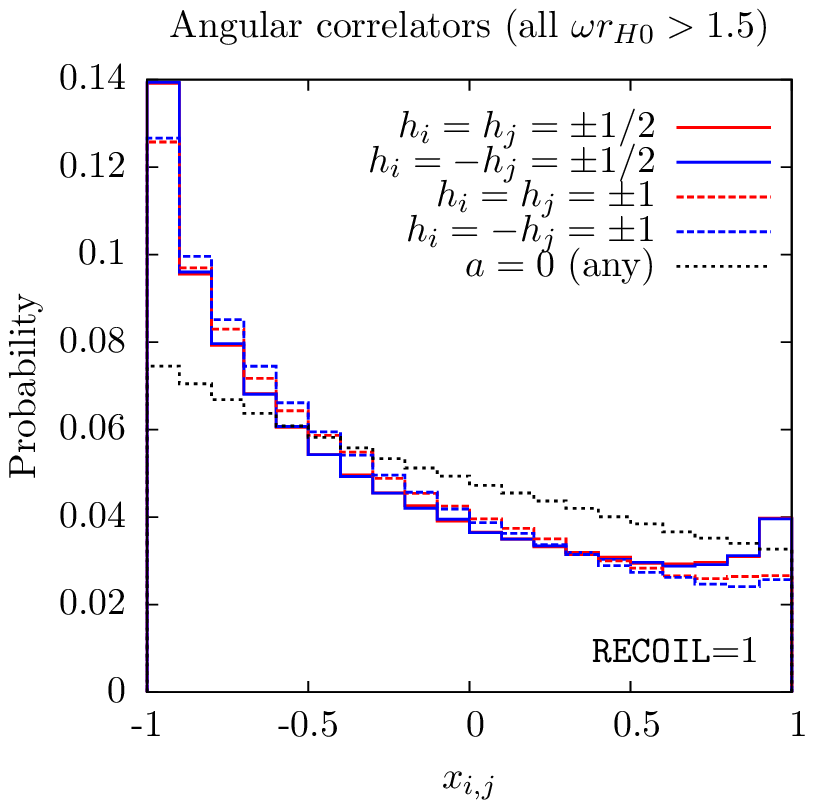} \vspace{1mm}\\
  \includegraphics[clip=true,trim=23 0 2 0,scale=0.6]{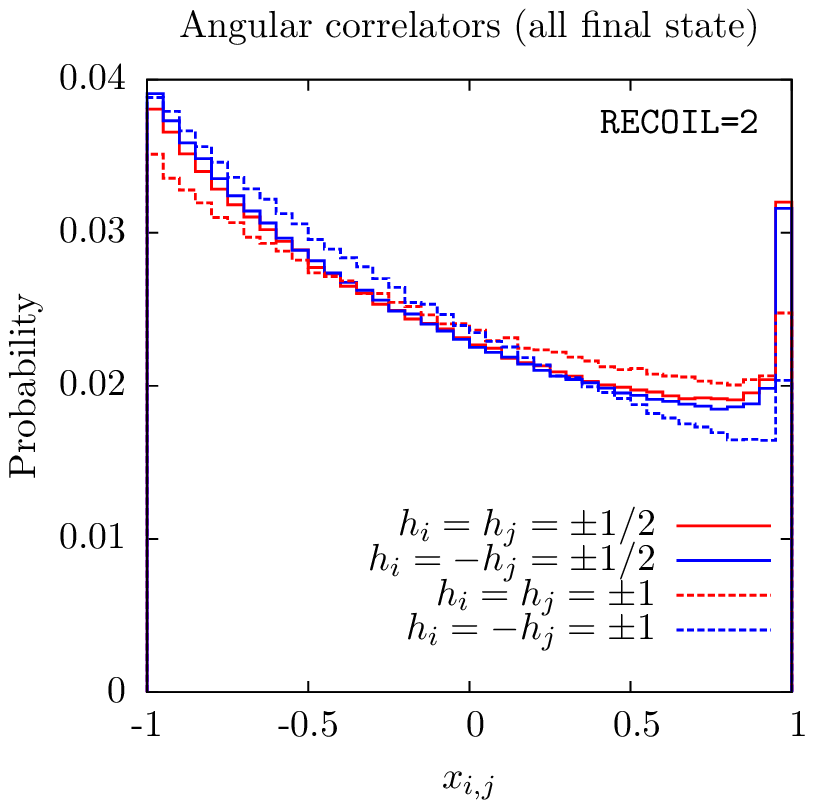}
  \includegraphics[clip=true,trim=20 0 2 0,scale=0.6]{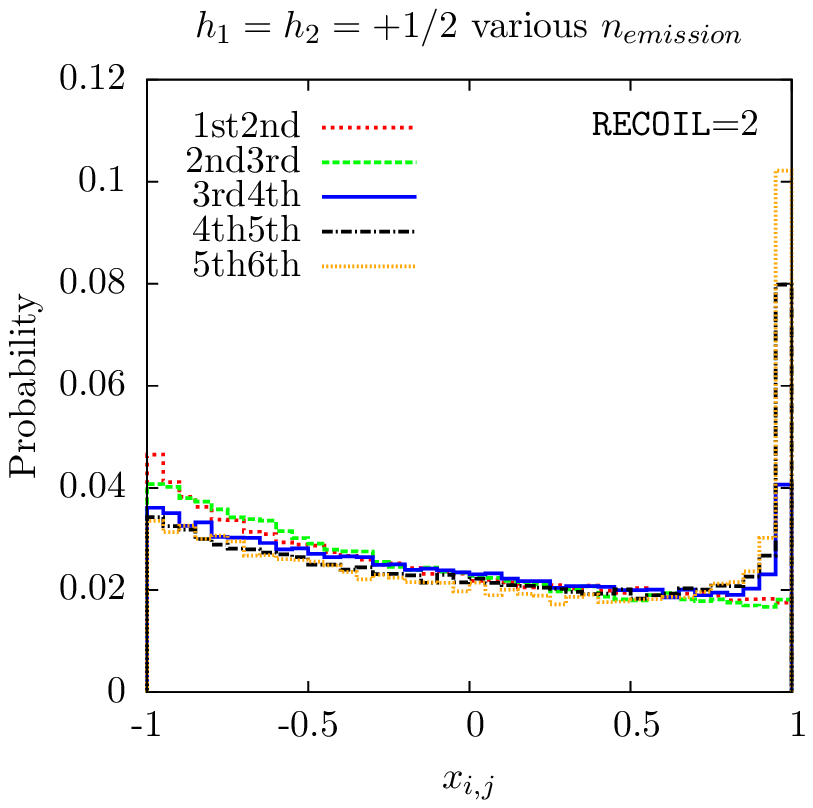}
  \includegraphics[clip=true,trim=20 0 2 0,scale=0.6]{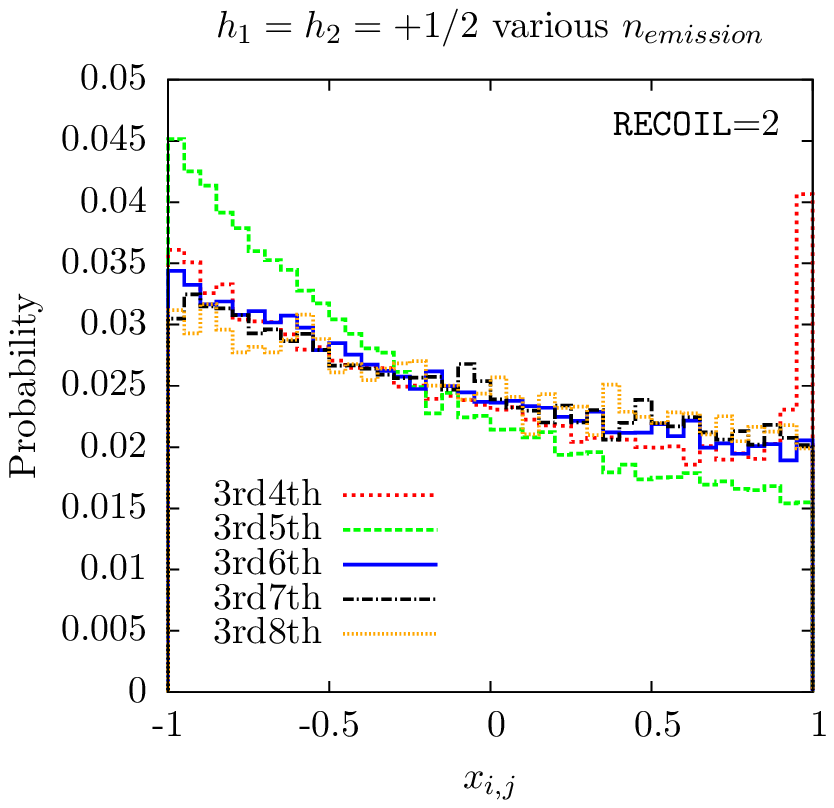}
\caption{\label{fig:MC_correlators} {\em Parton level angular correlators evaluated in the frame of the initial black hole.} The sample used was the same as in figure~\ref{fig:Jrec_spins}. The top left plot shows correlators between any two final state fermions or any two final state vector bosons for some helicity combinations. The correlator for any two particles when rotation is off is shown for comparison. The top centre and top right plots show the same distributions using various pairs of particles according to their ordering during the evaporation. The central row of plots shows the same distributions using specific intervals of energy (compared to the horizon radius for the initial black hole -- $r_{H0}$). Twice the bin size was used to compensate for the lower statistics. The bottom row of plots contains the same plots as the top row, but with a different recoil model \texttt{RECOIL}=2.}
\end{figure} 
Figure~\ref{fig:MC_correlators} shows the correlators defined in~\eqref{eq:x_def} for various helicity combinations. The top left plot shows the distribution for same helicity (red) and opposite helicity (blue) fermions (solid lines) and vector particles (dashed lines) and the non-rotating case (black) for comparison. The black curve is not constant due to the recoil of the black hole between emissions. The recoil tends to make different pairs of emissions more back to back (especially subsequent emissions) which causes the rise at $x_{i,j}\sim -1$ and the fall off at $x_{i,j}\sim 1$.

We can see (particularly for vector particles) that the asymmetry predicted in figure~\ref{fig:th_correlators} persists, with an enhancement at $x_{i,j}=-1$ for opposite helicity correlators, compared to the same helicity correlators. It is also clear that the effect of the recoil between emissions is larger in the rotating case. This is because the spectrum is harder, hence the relative boost when the black hole recoils is also larger. Additionally, in the rotating case, we have a forward peak at $x_{i,j}=1$. This is due to the possibility of having an accumulated boost in a certain direction for the particles emitted later in the decay, so pairs of those particles will tend to be more collinear. Since in the rotating case the multiplicity of the event is reduced and the spectrum is harder, this effect tends to be important (practically all events end up with a large accumulated boost). This is verified in the top centre and top right plots where for the first pair in the beginning of the evaporation (top centre plot) no forward peak is observed, whereas for later consecutive pairs, the forward peak tends to increase rapidly. For pairs which are increasingly separated in the order of emission, the forward peak also disappears (top right plot). The reason why no forward peak appears in the non-rotating case can be justified as follows. For events with multiplicity $N$, the number of consecutive pairs which can contribute to a forward peak is $N-2$, whereas the total number of pairs is $N(N-1)/2$. So the fraction of pairs contributing to a forward peak is at most $2(N-2)/(N(N-1))$. Typically, when rotation is turned off, the multiplicity increases from $5\sim 8$ to  $10\sim 15$ so the fraction of pairs contributing to the peak is reduced roughly by a factor of $\sim 2$. Furthermore, a larger multiplicity means a smaller magnitude for the boost in the recoil by another factor of $\sim 3$ (since the same energy is distributed among more particles which are softer). So overall, we have a suppression factor of at least $\sim 6$. This explains the absence of the forward peak for the non-rotating sample.

The central row of plots shows that by selecting particular ranges of energy we recover the strong asymmetry for low energy vector bosons (left and centre plots). For high energy particles (right), all helicities are equivalent. This is because at high energies all the angular spectra become equatorial regardless of the spin.  

The bottom row plots are the same as in the first row, except that \texttt{RECOIL}=2 was used (see section~3.3 of~\cite{Frost:2009cf}). Similar conclusions are obtained with this option (particularly for the ranges of energy selected in the middle row of plots). The only difference is that the forward peak due to the recoil is somewhat sharper and the backward effect is smaller (see explanation in appendix~\ref{app:recoil_effect}).

\section{Conclusions}
\label{sec:Conclusions}  

In this article, we have performed the first parton level study of angular asymmetries in semi-classical black hole events at the LHC. 

In the first part, we have done the first complete analysis of the geometrical optics discs for our background, which provides information on the Hawking fluxes at high energies, and a picture of the background as seen by an asymptotic observer. The calculation was shown to match with some properties of the full Hawking fluxes so it provides an alternative to model this asymptotic regime.

Phenomenologically, we have devised several strategies to extract angular information from the events. In the first type (axis reconstruction) though the reconstructed axis correlates with the true axis, there is a large fraction of wrong reconstructions due to recoil. This means that an angular distribution of the momentum of a particle with respect to such axis, is likely not to present a strong correlation. After observing this,  we have turned to angular correlators between pairs of particles, which do not depend on axis reconstruction and explore much better the angular asymmetries of the events. We have devised a set of energy cuts which are in principle easily generalised to a realistic experimental situation by using the energy scale of the process (which determines $r_{H0}$). 

We should note that the plots in figure~\ref{fig:MC_correlators} were produced assuming knowledge of the helicities of the outgoing particles. This is usually not a direct observable at hadron colliders. However the decay modes of the the $W$ and $Z$ bosons are dependent on the helicity of the intermediate states. For example in the decay $W^{-}\rightarrow \ell^{-}\bar{\nu_\ell}$ the charged lepton tends to be collinear with the $W^{-}$ for negative helicity, and anti-collinear for positive helicity. For $Z$ decays a similar argument holds. The only disadvantage of $W$ decays is the neutrino which makes re-construction harder, whereas for $Z$ decays the disadvantage is the mixture of left-handed and right-handed couplings. Finally, gluons, being vector particles should also exhibit similar asymmetries, so cutting on lower energy jets could potentially produce similar asymmetries, even without helicity knowledge.

To conclude, our analysis shows that there is an effect which in principle can be extracted from the data at the experimental level should black hole candidates be found in the future runs at the LHC. A full detector level analysis, which is beyond the scope of this work, will tell us how to extract in detail these interesting effects.

\section*{Acknowledgements}
 I thank colleagues in the Cambridge SUSY working group for discussions at the Cavendish Laboratory, while this work was produced. The author is supported by the FCT grant SFRH/BPD/69971/2010. This work is also supported by the grants CERN/FP/116341/2010, PEst-C/CTM/LA0025/2011 and NRHEP–295189FP7-PEOPLE-2011-IRSES.

\section*{Appendices}
\appendix

\section{Critical impact parameters}
\label{app:Critical}
 In this appendix we describe how to obtain the critical impact parameter discussed in the main text. When $Q=0$
\begin{eqnarray}\label{ABC}
A&=&r^{n+2}+r^na^2\left(1-\cos^2{\zeta}\sin^2{\vartheta}\right)-(1+a^2)r \nonumber\\
B&=&-a\cos{\zeta}\sin{\vartheta}\left(1+a^2\right)r \nonumber\\
C&=&r^{n+4}(1-\nu^2)+r^{n+2}a^2(1-\nu^2+\cos^2{\vartheta})+r^na^4\cos^2{\vartheta}+ \nonumber\\
 &&+r^3\nu^2\left(1+a^2\right)+ra^2(1+a^2)\sin^2{\vartheta}
\end{eqnarray}
where $C\geq0$ since $\nu\leq 1$. For a fixed $b$, it can be shown that $A$ has exactly one zero in the domain $r>0$ so it changes sign only once\footnote{This is done by looking at the sign of the function at infinitesimal $r$ and at infinity, using continuity and the positivity of $d^2A/dr^2>0$ for $r>0$.}. So $A$ starts from positive values at infinity, decreases, goes through zero and takes negative values inwards. Since the signs of $B$ and $C$ are fixed we only need to analyse two regions. 

Let's start with the outermost region. We want to determine the impact parameters for which the particle is not able to penetrate completely through the region of $A\geq 0$, thus being scattered back at some point, or equivalently the range of parameters $b_{min}<b<b_{max}$ for which absorption is guaranteed. Then it is certain that any particle with impact parameters within this range will reach the second innermost region. For any solution of the radial equation we must ensure  that $\mathcal{R}\geq 0$ to allow the particle through the region. As a function of $b$, $\mathcal{R}$ has two zeros
\begin{equation}
b_\pm=\dfrac{B\pm\sqrt{B^2+AC}}{A}
\end{equation}
Since, $AC$ is positive, there is a negative and a positive root. Thus $\mathcal{R}$ is a parabola with a maximum. $\mathcal{R}$ is positive only for $0<b<b_+$. If we take $b>b_{max}\equiv\min_{r}\left\{b_+\right\}$, where the minimisation is over the radial region we are considering, there will always be a point where $\mathcal{R}$ goes through zero and changes sign. This means that the particle is scattered back at that point. Similarly, in the complementary case, there will never be such a zero so the particle reaches the second interior radial region. Thus we have obtained the first upper bound.

Regarding the remaining interior radial region the situation is a bit more complicated. In that case, $A<0$ so the roots are
\begin{equation}
b_\pm=\dfrac{-B\mp\sqrt{B^2-|A|C}}{|A|} \; .
\end{equation}
For $B>0$ (or equivalently $b_z<0$, see \eqref{reduced_vars}), the previous equation has no positive real roots, so $\mathcal{R}>0$ and any particle reaching the region is absorbed. Thus, for this sign of $B$, the absorptive disc is defined by the interval obtained in the outer region. For $B\leq0$, we can have two real positive roots. Those are
\begin{equation}
b_\mp=\dfrac{|B|\mp\sqrt{B^2-|A|C}}{|A|}
\end{equation}
so $\mathcal{R}$ is a parabola with a minimum and two zeros. If $b$ takes any value below $\min_r\{b_{-}\}$ (again the minimisation is in the radial region we are considering), there is no zero, because any $r$ has associated critical $b$'s which are necessarily larger. The same occurs for $b$ above $\max_{r}\{b_{+}\}$, since all the $b$'s that allow a zero are smaller. Conversely, for $b$ between the last two values, there is  always a point where $\mathcal{R}$ goes through zero and becomes negative, so the particle is scattered back. Therefore, the impact parameters for absorption must be smaller than $\min_{r}\left\{b_{-}\right\}$ or larger than $\max_{r}\left\{b_{+}\right\}$. However, note that at $r=r_{A}$, ($r_{A}$ defined such that $A(r_{A})=0$) $b_{+}$ diverges, so there is no finite $b>\max_{r}\left\{b_{+}\right\}\rightarrow+\infty$.

\subsection{Recurrence relations}
\label{app:Recurrence}
Since $r_c,b_c$ are the minimiser and minimum when $a\neq0$, $r_c$ must obey
\begin{equation}\label{cond_xmin}
\left.\dfrac{\partial b}{\partial r}\right|_{r=r_c}=0 \; ,
\end{equation}
so that
\begin{equation}\label{perturb_bc}
b_c=\left.b\right|_{r=r_c} \; .
\end{equation}
Note that from now on, $b$ is given by the expression to be minimised in~\eqref{def_min_impact} and all other parameters $\zeta,\vartheta,n$  are omitted.
Expanding~\eqref{cond_xmin} in powers of $a$ and imposing each coefficient to vanish independently, we obtain conditions on the $r_p$ coefficients (see equation~\eqref{def_x_min})
\begin{equation}\label{x_p_conditions}
\left.\dfrac{d^m}{da^m}\left(\left.\dfrac{\partial b}{\partial r}\right|_{r=r_c}\right)\right|_{a=0}=0 \; .
\end{equation}
One can show that the total derivatives can be expanded in the following manner
\begin{equation}\label{eq:deriv_expand}
\dfrac{d^m}{da^m} =\displaystyle{\sum_{k=0}^1{\left(\dfrac{dr_c}{da}\right)^k\dfrac{d^{m-1}}{da^{m-1}}\dfrac{\partial}{\partial a^{1-k} \partial r_c^{k}}}+\dfrac{d^mr_c}{da^m}\dfrac{\partial}{\partial r_c}} \; .
\end{equation}
By successively iterating~\eqref{eq:deriv_expand}, one can find a general expression which is proved by induction
\begin{multline}\label{da_expand}
\dfrac{d^m}{da^m} =\sum_{q=0}^{m-1}{\dfrac{(m-1)!}{(m-1-q)!q!}\left(\dfrac{dr_c}{da}\right)^q\dfrac{\partial^m}{\partial a^{m-q} \partial r_c^{q}}}+\\
+\sum_{q=0}^{m-1}\dfrac{d^{m-q}r_c}{da^{m-q}}\sum_{k=0}^q{\dfrac{q!}{(q-k)!k!}\left(\dfrac{dr_c}{da}\right)^k\dfrac{\partial^{q+1}}{\partial a^{q-k} \partial r_c^{k+1}}} \; .
\end{multline}
Using~\eqref{da_expand} in~\eqref{x_p_conditions}, relabelling $r_c\rightarrow r$ in the partial derivatives, and defining
\begin{equation}
\beta_{i,j}\equiv\left.\dfrac{\partial^{i+j}b}{\partial a^{i} \partial r^{j}}\right|_{a=0,r=r_0}
\end{equation}
gives the recursion relations ($m>1$) for all the corrections
\begin{equation}
\begin{array}{rcl}
r_1\beta_{0,2}&=&-\beta_{1,1} \\
r_m\beta_{0,2}&=&-\beta_{m,1}-\displaystyle{\sum_{q=1}^{m-1}\left[\dbinom{m-1}{q}r_1^q\beta_{m-q,q+1}+r_{m-q}\sum_{k=0}^{q}r_1^k\dbinom{q}{k}\beta_{q-k,k+2}\right]} \; .
\end{array}
\end{equation}
The first correction is
\begin{equation}
r_1=-\dfrac{2\cos{\zeta}\sin{\vartheta}}{\sqrt{(n+1)(n+3)}} \; .
\end{equation}
Furthermore, a similar expansion for $b_c$ can be found by using \eqref{da_expand}. Define
\begin{equation}
b_c\equiv\sum_{m=0}^\infty \frac{b_m}{m!}a^m
\end{equation}
where
\begin{equation}
b_{m+1}\equiv\left.\dfrac{d^{m+1}b_c}{da^{m+1}}\right|_{a=0}=\left[\dfrac{d^{m}}{da^{m}}\left(\dfrac{\partial}{\partial{a}}+\dfrac{dr_c}{da}\dfrac{\partial}{\partial r_c}\right)b_c\right]_{a=0}=\left[\dfrac{d^{m}}{da^{m}}\dfrac{\partial b_c}{\partial{a}}\right]_{a=0} \; .
\end{equation}
The last step follows from \eqref{x_p_conditions}. The result is ($m\geq1$)
\begin{equation}
\begin{array}{rcl}
b_{1}&=&\beta_{1,0} \vspace{3mm}\\
b_{m+1}&=&\displaystyle{\sum_{q=0}^{m-1}\left[\dbinom{m-1}{q}r_1^q\beta_{m+1-q,q}+r_{m-q}\sum_{k=0}^{q}r_1^k\dbinom{q}{k}\beta_{q+1-k,k+1}\right]} \; .
\end{array}
\end{equation}
The first correction is actually independent of the perturbation in $r$
\begin{equation}
b_1=-\dfrac{2\cos{\zeta}\sin\vartheta}{n+1} \; .
\end{equation}

\section{Eternal black hole angular correlators}
\label{sec:ditribs_eternal}
In this section we obtain the angular correlators on an eternal black hole background, used in the main text.
We define the probability density function for the correlator $x_{i,j}$ (up to a normalisation constant)
\begin{equation}
\label{eq:def_rhoxij}
\rho(x_{i,j}) \propto \int dx_i dx_j d\phi_id\phi_j\rho_i(x_i)\rho_i(x_j)\delta\left(x_{i,j}-\sqrt{1-x_i^2}\sqrt{1-x_j^2}\cos(\phi_i-\phi_j)-x_ix_j\right)
\end{equation} 
where we have defined the spatial momenta of particle $i$ (or $j$)
\begin{equation}
\mathbf{p}_i=\left(\sqrt{1-x_i^2}\cos\phi_i,\sqrt{1-x_i^2}\sin\phi_i,x_i\right)|\mathbf{p}|
\end{equation}
and 
\begin{equation}
\rho_i(x_i) = \sum_{K=0}^{+\infty}\int_0^{+\infty}d\omega \frac{\mathbb{T}^{(D)}_{k}(x, a_*)}{\exp(\tilde{\omega}r_H/\tau_H) \pm 1} |S_{k}(a\omega,x_i)|^2
\end{equation} 
is the probability density of having a particle of type $i$ (with any energy) emitted with direction $x_i$ with respect to the angular momentum axis. Due to the azimuthal symmetry, the distribution is uniform in the $\phi_i$ direction. This can be written in a more convenient form by using the definition
\begin{equation}
C_{h,j,m,a_{*},D}(\omega r_H)=\int_0^{\omega r_H}\mathrm{d}x\frac{1}{\exp(\tilde{x}/\tau_H) \pm 1} \mathbb{T}^{(D)}_{k}(x, a_*)   
\end{equation} 
to obtain
\begin{equation}
\rho_i(x_i) =\sum_{K=0}^{+\infty}\int_0^{+\infty}d\omega \dfrac{dC_{h,K,a_{*},D}(\omega r_H)}{d\omega} |S_{k}(a\omega,x_i)|^2 \; .
\end{equation}
Then making the change of variable $y=f_{K}(\omega)\equiv C_{h,K,a_{*},D}(\omega r_H)/C_{h,K,a_{*},D}(+\infty)$
\begin{eqnarray}
\rho_i(x_i)&=&\sum_{K=0}^{+\infty}\int_0^{1} dy\, C_{h,K,a_{*},D}(+\infty) |S_{k}(a f^{(-1)}_K(y),x_i)|^2 \nonumber\\
&=&\int_0^{1} dy\sum_{K=0}^{+\infty} C_{h,K,a_{*},D}(+\infty) |S_{k}(a f^{(-1)}_K(y),x_i)|^2
\end{eqnarray} 
where $f^{(-1)}$ is the inverse function (not $1/f$). Now if we define
\begin{eqnarray}
C_i(x_i)&\equiv&\int_{-1}^{x_i}dx\, \rho_i(x) \nonumber\\
&=&\int_0^{1} dy\sum_{K=0}^{+\infty} C_{h,K,a_{*},D}(+\infty)\int_{-1}^{x_i}dx |S_{k}(a f^{(-1)}_K(y),x)|^2 \\
\Rightarrow \rho_i(x_i)&=&\dfrac{dC_i(x_i)}{dx_i} \; .
\end{eqnarray}
Going back to equation~\eqref{eq:def_rhoxij} we perform the changes of variables\begin{eqnarray}
\Phi&=&\phi_i+\phi_j \\
\phi&=&\phi_i-\phi_j \\
w_i&=&g_i(x_i)=C_i(x_i)/C_i(1)
\end{eqnarray}
to obtain
\begin{eqnarray}
\rho(x_{i,j}) &\propto& \int_{0}^1\int_0^1\int_0^{4\pi}\int_{-2\pi}^{2\pi} dw_i \, dw_j \, d\Phi \,d\phi \, \delta\left(x_{i,j}-\sqrt{1-x_i^2}\sqrt{1-x_j^2}\cos\phi-x_ix_j\right)\nonumber \\
\Rightarrow\rho(x_{i,j})&= & \dfrac{1}{2\pi}\int_{0}^1\int_0^1\int_{0}^{2\pi} dw_i \, dw_j \,d\phi \, \delta\left(x_{i,j}-\sqrt{1-x_i^2}\sqrt{1-x_j^2}\cos\phi-x_ix_j\right) 
\end{eqnarray}
where $x_i=g^{(-1)}(w_i)$ and we have normalised the distribution. 
The histogram for $\rho(x_{i,j})$ is obtained by generating the phase space $w_i,w_j,\phi$ uniformly and adding a unit weight to the bin for the corresponding
\begin{equation}
x_{i,j}=\sqrt{1-x_i^2}\sqrt{1-x_j^2}\cos\phi+x_ix_j \; .
\end{equation}

\section{The effect of recoil}
\label{app:recoil_effect}
In this section we explain the backward and forward peaks found in figure~\ref{fig:MC_correlators}, using the kinematics of the two recoiling options in \texttt{CHARYBDIS2}. Using the description in~\cite{Frost:2009cf}, it can be shown that the mass reduction for a particle with a small energy $E \ll M$ (i.e. in the beginning of the evaporation) in the frame of the initial black hole is
\begin{equation}
M_{final}=\left\{\begin{array}{lc}
M-E\left(1+\dfrac{E}{2M}\right)+\ldots & , \texttt{RECOIL}=1 \vspace{2mm}\\ 
M-E  & , \texttt{RECOIL}=2 
\end{array} \right. \; .
\end{equation} 
 This explains the larger backward effect (at $x=-1$) in the top plots when \texttt{RECOIL}=1, since in that case the black hole mass reduction is a bit larger in the beginning of the evaporation, enhancing the effect of the back to back recoil. The sharper forward peak when \texttt{RECOIL=2} can be explained by expressing $E$ in terms of the selected Hawking energy $\omega$ (neglecting particle mass)
\begin{equation}
\dfrac{E}{M}=\left\{\begin{array}{lcc}
\dfrac{\omega}{M} &, \dfrac{\omega}{M} \in [0,\frac{1}{2}] & , \texttt{RECOIL}=1 \vspace{2mm}\\ 
\dfrac{\omega}{M}\left(1-\dfrac{1}{2}\dfrac{\omega}{M}\right) & , \dfrac{\omega}{M} \in [0,1] & , \texttt{RECOIL}=2 
\end{array} \right. \; .
\end{equation}
\begin{figure}
\centering
  \includegraphics[clip=true,scale=0.8]{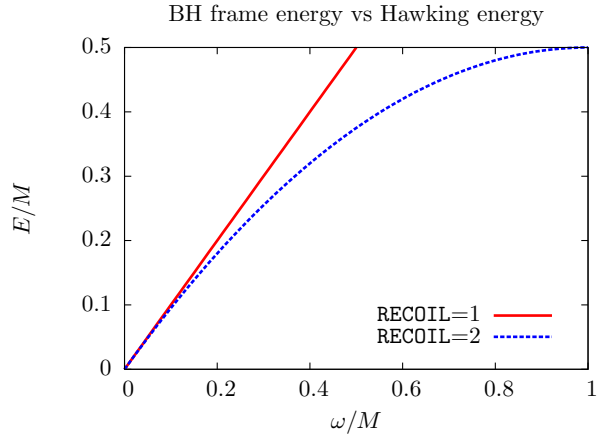}
\caption{\label{fig:recoil_kin} {\em Energy of the particle emitted:} The energy of the particle ($E/M$) in the frame of the black hole is shown as a function of the energy selected from the Hawking spectrum ($\omega/M$).}
\end{figure} 
It is easy to see (figure~\ref{fig:recoil_kin}) that at high energies $E$ close to the kinematic limit (which are more important in the last part of the evaporation), the range of energies $\omega$ contributing to a range of energies $E$ is always much wider for \texttt{RECOIL}=2. So there will be more hard particles selected at the end of the evaporation for the latter, contributing to the accumulated boost and hence the forward peak. 

\bibliography{references}
\bibliographystyle{JHEP}

\end{document}